\documentclass[%
preprint,
amsmath,amssymb,
aps,
]{revtex4-2}
\usepackage{graphicx}
\usepackage{color}
\usepackage{booktabs}

\begin{document}


\title{Long-range order Bragg scattering and its effect on the dynamic response of a Penrose-like phononic crystal plate}


\author{Domenico Tallarico}
\email[]{domenico.tallarico@empa.ch}
\affiliation{EMPA, Laboratory for Acoustics/ Noise Control, \"{U}berlandstrasse 129, 8600 D\"{u}bendorf, Switzerland}
\author{Andrea Bergamini}
\affiliation{EMPA, Laboratory for Acoustics/ Noise Control, \"{U}berlandstrasse 129, 8600 D\"{u}bendorf, Switzerland}
\author{Bart Van Damme}
\affiliation{EMPA, Laboratory for Acoustics/ Noise Control, \"{U}berlandstrasse 129, 8600 D\"{u}bendorf, Switzerland}


\date{\today}

\begin{abstract}
In this article, we present scattering and localization phenomena in a thin elastic plate comprising an aperiodic arrangement of scatterers. By analysing the form factor of the scattering cluster, we sample the reciprocal space which shows strong scattering points  associated with non trivial dispersion. Wide frequency regimes with very different dynamic responses  are identified: isotropic wave dispersion at low frequency, and an attenuating regime with strong localization effects at higher frequencies.  Illustrative comparisons are drawn with a periodic counterpart having the same density of scatterers. The novel findings are corroborated by analytical estimates, numerical finite-element simulations and vibrometric experiments. The results are relevant for the research community interested in extending phononic crystal phenomena to lower frequencies. 
\end{abstract}


\maketitle

\section{Introduction}
Quasi-crystals are ordered tessellations of space without translational symmetry. The first natural quasi-crystal   featuring icosahedral point group symmetry - inconsistent with lattice translations - was reported in the early 80s by Shechtman \emph{et al.} \cite{shechtman1984metallic}. At the same time, solid state physics have inspired theoretical and experimental studies into metamaterials, man-made structures whose non-trivial dispersive behaviour has found applications into many branches of classical physics, such as photonics, acoustics, continuum mechanics and thermodynamics, providing solutions to plenty of engineering problems. At the basis of metamaterials' pervasive success, there is the so called Bloch-Floquet theorem which guarantees that the fundamental solutions of periodically modulated structures can be  traced back to the solution of relatively cheap eigenvalue problems. 

Locally-resonant metamaterials  are particularly successful for structural dynamics due to the fact that their constituents are active in the deep subwavelength regime. The resonating scatterers are placed less than half a wavelength apart, resulting in a high imaginary part of the wave number and thus rapidly decaying elastic waves. However, such sub-wavelength behaviour is accompanied by narrow-band effects. On the other hand, the frequency bands  for dispersive effects  in phononic crystals are inherently linked to the size of the unit cell, while being wider compared to  locally-resonant  counterparts.   

Limited attention has been given so far to aperiodic phononic quasi-crystals. One-dimensional examples have been reported with a focus on longitudinal elastic wave dynamics in Fibonacci-like rods \cite{gei2020phononic}; flexural wave dynamics in beams with their thickness modulated according to incommensurable periodicities \cite{van2021bending}; topological band gaps in beams with aperiodically-modulated local resonances \cite{rosa2021exploring}. In two-dimensions, Beli \emph{et al.} \cite{beli2021mechanics} introduced by-design decoration of plates featuring point-wise Bragg peaks with 8-, 10- and 14- fold rotational symmetry. The authors also report highly isotropic wave propagation within the aperiodic composites. The results are extended in \cite{beli2022wave} to account for wave beaming. 

\begin{figure}[h!]
	\centering
	\includegraphics[width=0.445\textwidth]{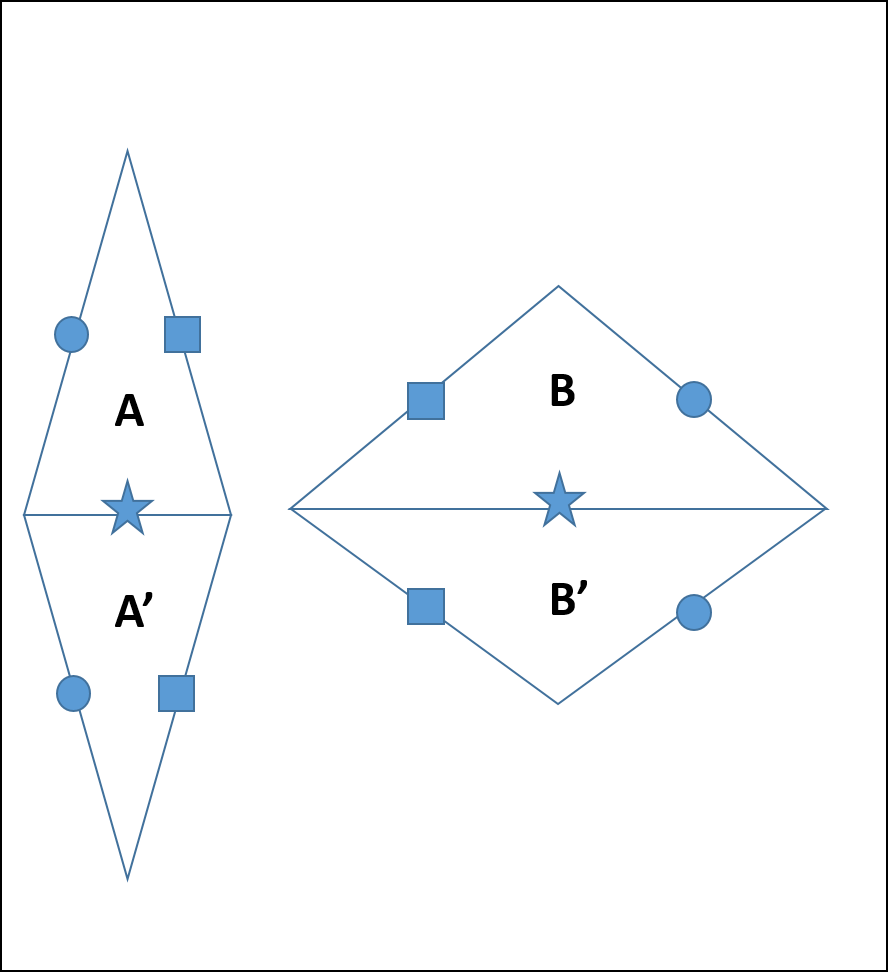}\hfill	
	\includegraphics[width=0.5\textwidth]{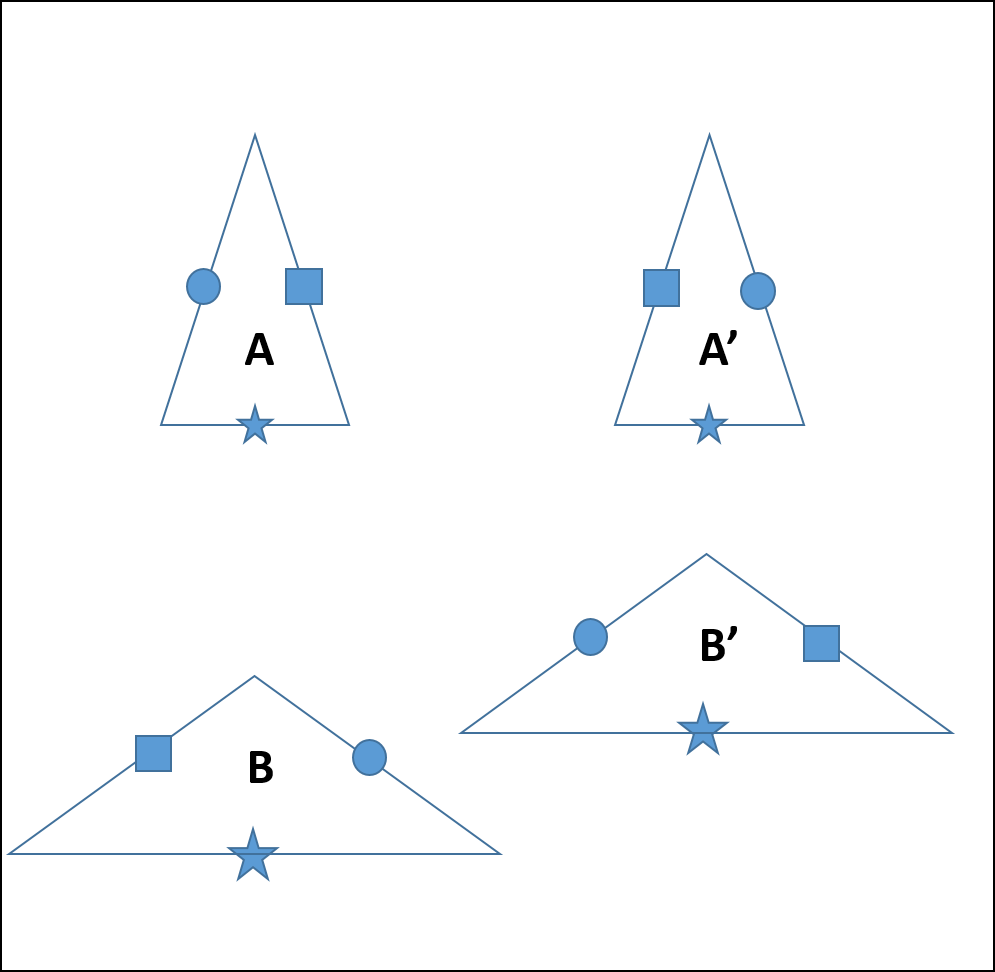}\\
	\vspace{1cm}
	\includegraphics[width=1\textwidth]{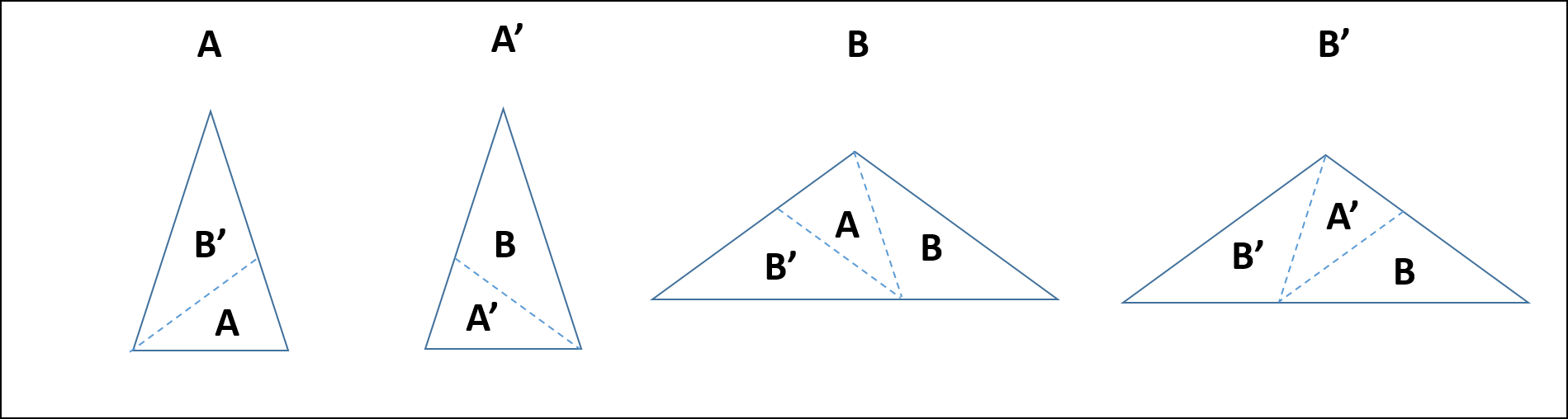}
	\begin{picture}(0,0)(0,0)
	\put(-130,158) {$(a)$}
	\put(100,158) {$(b)$}
	\put(15,0) {$(c)$}
	\end{picture}
	\caption{\label{fig:geom} Overview of the construction rules for P3 Penrose tiling, comprising a thin and a thick rhombus, whose side lengths are $\ell_0$.}
\end{figure}
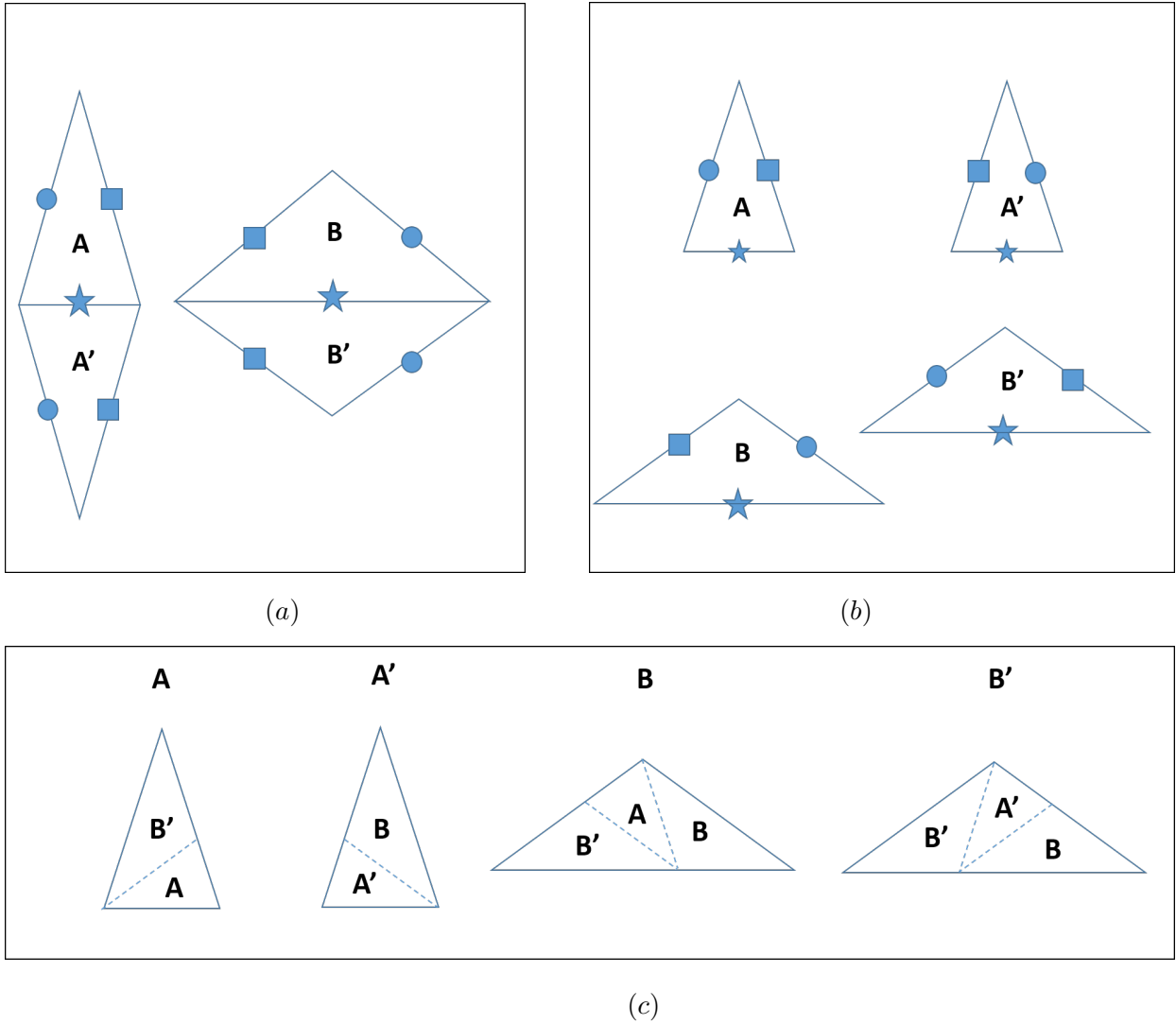

In the present article we give attention to the effect of long-range order in a P3 Penrose quasi-crystal \cite{penrose1979class} on the dynamic response of finite clusters of resonators (either embedded into an infinite plate or hosted by a finite plate). Such quasi-crystal can be constructed by a so-called deflation/inflation tiling, illustrated in Fig. \ref{fig:geom}. We give a concise introduction and we refer to Ref. \cite{grimm2021highly} for a more detailed discussion. Each rhombus  (panel (a)) can be divided into two isosceles triangles. Each pair of triangles (A,A' and B,B', see Fig. \ref{fig:geom}(b)) differ by the way equal sides are decorated (either a circle or a square appear on the right hand side of the top vertex). Rhombuses are obtained by joining the pair elements along the ``starred'' side. Adjacent rhombuses are added by joining triangles along equal sides, i.e. sharing the same length and symbol, a process referred to as \emph{inflation}.  Each triangle is decomposed according to the rules outlined in Fig. \ref{fig:geom}(c), where we show only the first level of decomposition. An arbitrary level of decomposition can be obtained by applying the rules to the inner triangles, giving rise to a so called \emph{deflation} rule. In this paper we decompose the BB'  rhombus (Fig. \ref{fig:geom}(a)). The $n^{\rm th}$ level of decomposition induces the rhombuses' length side $\ell_n = (2 \sin(\vartheta_0))^{-n} \ell_0$, with $\vartheta_0=54~{\rm deg}$. A Matlab implementation \cite{eddins2021matlab} of the tessellation has been used to  generate the aperiodic distribution of scatterers.

 We note that aperiodic tilings of the plane can also be introduced via \emph{cut and project} techniques, \emph{i.e.} by projecting part of the four-dimensional periodic tiling. Since the number of possible symmetries increases with dimensions, the projected  aperiodic  counterparts inherit  possible underling five-, eight-, ten- or twelve-fold symmetry. Formal results exist showing the point nature of the $n$-fold  diffraction pattern of several cut and project lattices. While establishing such formal results for inflation/deflation structures is notoriously more complicated, assuming that the scattering potential is a Dirac comb leads to a closed-form expression  for structure factor and allows a visualization of the diffraction pattern (see section \eqref{sec:structure_factor}).

The main focus is in singling out the effect of Bragg-scattering of flexural waves in a thin plate due to the presence of a P3 Penrose cluster of scatterers.  After fixing the decomposition level, we associate an elastic stud to  the centre of mass of each resulting triangle (of type A,A',B or B'). Endowing rigid scatterers with rotational degrees of freedom and the corresponding inertia allows a realistic representation of the experimental setup, adding complexity to the case of point-mass scatterers driven only by the flexural field, as described in Sec. \ref{sec:results}.  In this article, we focus on the   distribution of scatterers  represented in Fig. \ref{fig:distro}. 

The number of scatterers at the $i^{\rm th}$ decomposition levels, can be evaluated by accounting for the total number  of ${\rm A}$ and ${\rm A}'$ triangles ($N_{\rm A}$), and  the total number of ${\rm B}$ and ${\rm B}'$ triangles ($N_{\rm B}$). Specifically, from Fig. \ref{fig:geom}(c) the  following recursive relations  for the total number of triangles can be inferred 
\begin{eqnarray}\label{eq:recurrence}
&N_{\rm A}^{(i)} = N_{\rm A}^{(i-1)}+N_{\rm B}^{(i-1)},\nonumber\\
&N_{\rm B}^{(i)} = N_{\rm A}^{(i-1)}+2 N_{\rm B}^{(i-1)},~~~~ i = 1,\cdots,n,
\end{eqnarray}
which allows to calculate $N_A=N_A^{(n)}$ and $N_B=N_B^{(n)}$, and therefore $N_s=N_A+N_B$, given a seed number of initial triangles $N_B^{(1)}$ and $N_A^{(1)}$. 

The article is organised as follows. In section \ref{sec:gov} we present the Penrose-like distribution of scatterers, as well as the governing equations, useful to describe the phononic crystal under consideration. In section \eqref{sec:results}, we present and discuss the results concerning both a finite cluster of non-resonant scatterers hosted by an infinite plate and a finite plate hosting the same arrangement of scatterers. Finally, in section \ref{sec:conclusions} we summarise the results and give our main conclusions.

\section{Structures and associated Governing equations \label{sec:gov}}
\subsection{Flexural waves in thin elastic plates}
Time-harmonic flexural waves in a thin plate comprising a cluster of resonators are governed by the partial differential equation
\begin{equation}\label{eq:governing-eq}
{\cal L} \psi({\bf r})=f(\beta) \sum_{j=1}^{N_s}  \delta ( {\bf r}-{\bf r}_j) \psi({\bf r}).
\end{equation} 
where $ \psi({\bf r})$ is the amplitude of the time-harmonic flexural field at the spatial coordinate ${\bf r}$ in the plane.  In  Eq. \eqref{eq:governing-eq} we  introduce the operator ${\cal L} =\Delta^2-\beta^4 $, with $\Delta^2$ the bi-harmonic operator and 
\begin{equation}\label{eq:equation}
\beta(\nu) = \left(\frac{(2\pi\nu)^2\rho_s h}{D}\right)^{1/4},
\end{equation}
the wave-number, where $\nu$ is the frequency, $\rho_s$ is the mass density and  $D=E_s h^3/(12(1-\nu_s^2))$ is the flexural rigidity, with $h$, $E_s$ and $\nu_s$ being the thickness, Young's modulus and Poisson's ratio, respectively.   

The Green's function associated with the operator ${\cal L} $ (i.e. the solution of ${\cal L}g_0({\bf r},{\bf r}')=\delta ( {\bf r}-{\bf r}')$ ) is \cite{watanabe2014integral}
\begin{equation}\label{eq:gf}
g_0({\bf r},{\bf r}';\beta)=\frac{i}{8 \beta^2} \left( H_0^{(1)}(\beta |{\bf r}-{\bf r}'|) + \frac{2 i}{\pi} K_0(\beta |{\bf r}-{\bf r}'|)\right), 
\end{equation} 
where $H_0^{(1)}(\beta |{\bf r}|)$ and $K_0(\beta |{\bf r}|)$ are the Hankel function of the first kind and the modified Bessel function of the second kind, respectively.   
The solution of the scattering problem to an incident field can be obtained by using the Korringa-Kohn-Rostoker (KKR) method which builds upon the knowledge of the Green's function of the homogenous operator and the scattering potential in the r.h.s. of Eq. \eqref{eq:governing-eq}. Details of the derivation are provided in \color{black}{Appendix \ref{sec:appendix}.} \color{black} We observe that $f(\beta)$ can be any function of frequency resulting from the physics of the considered resonator.  However, it should be pointed our that the model in Eq. \eqref{eq:governing-eq} is accurate only for scatterers exerting purely normal forces with respect to the plate and vanishing moments. This is the case for few but ideal cases, \emph{e.g.} sprung and unsprung point masses, which in real-life, can be only obtained via active control of vibrations or tackled with some special manufacturing design. In general, the inevitable finite extent of passive scatterers results in non-zero rotational inertia or even elastic deformations  which in turn couple with the rotation and in-plane fields of the Kirchhoff's plate. This calls for the need to consider separately the equations for the forces and moments, which are coupled by the boundary conditions between the plate and the scatterers. The construction of the solution for such coupled problem based on Green's function techniques only, is in general more complicated than that of a single scalar equation \eqref{eq:equation} and beyond the scope of this paper. The formulation presented in \cite{cai2016movable} extends the multiple scattering method with purely vertically-acting point mass  to the case where the scatterers rotational inertia is included. 

\subsection{Finite-element formulation \label{sec:fe_formulation}}

An alternative natural framework to accommodate the rotational inertia of the scatterers is provided by commercial finite-element (FE) solvers. In Ansys, the plate governing equations (Reissner-Mindlin plate theory \cite{elishakoff2017midlin})) are encoded into the SHELL281 element \footnote{ANSYS Element Reference, Release 2022 R2, 2022}.
 The formulation differs from the Kirchhoff's theory in Eq. \eqref{eq:equation} due to the account of first-order shear deformation effects and rotational inertia of the plate, which in the present paper are kept small by-design, the minimum wavelength being  $\lambda_{\rm min}>h/6$.
Nevertheless, in finite element routines, the kinematics of plates account for  displacement the rotation fields as independent variables,  allowing to consistently define boundary conditions of the plate's rotational and translational degrees of freedom with the rotations and translations of the rigid studs.  This consideration will be further illustrated in sections \ref{sec:periodic} and \ref{sec:experiment}. 
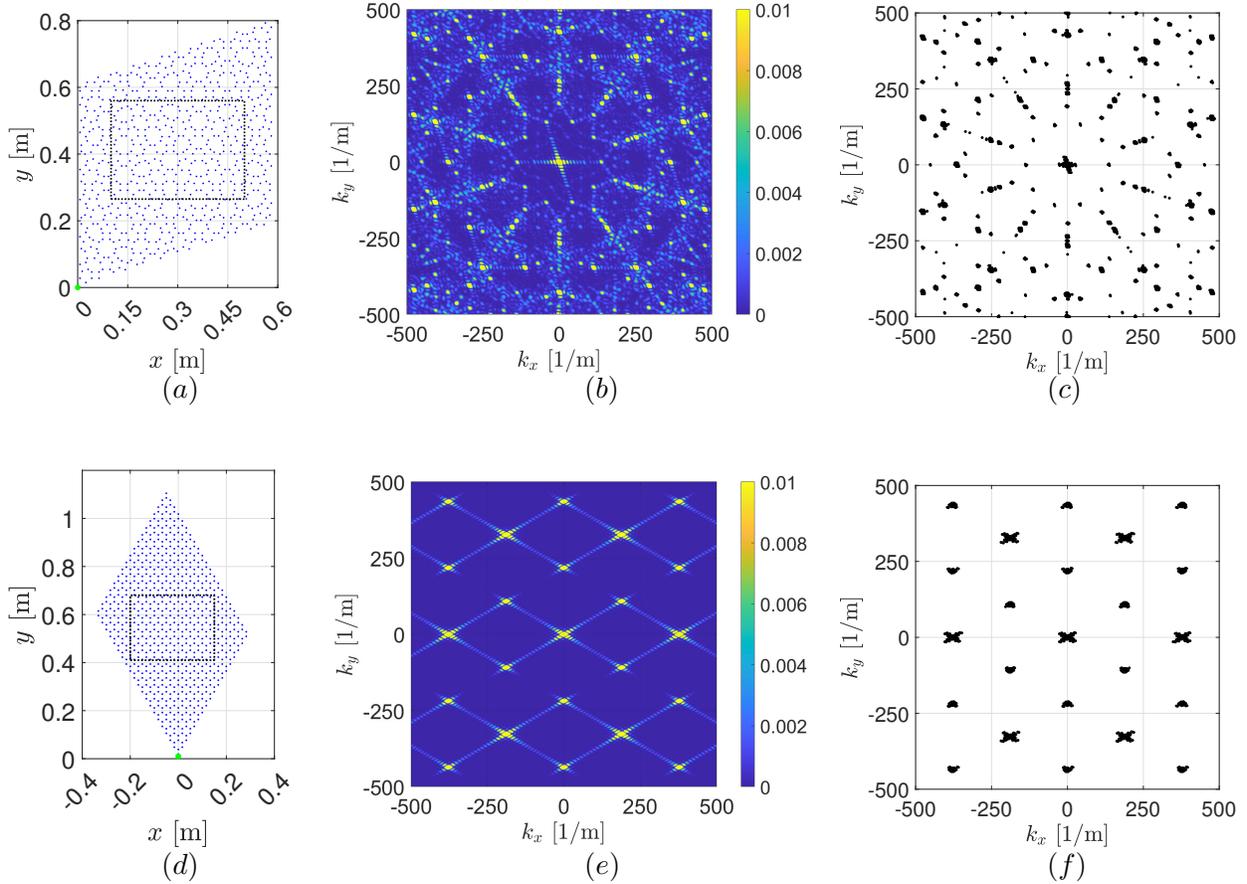
\begin{figure}[h!]
	\centering
	\includegraphics[
	trim = 0cm 0cm -1cm 0cm,clip,  
	width=0.25\textwidth]{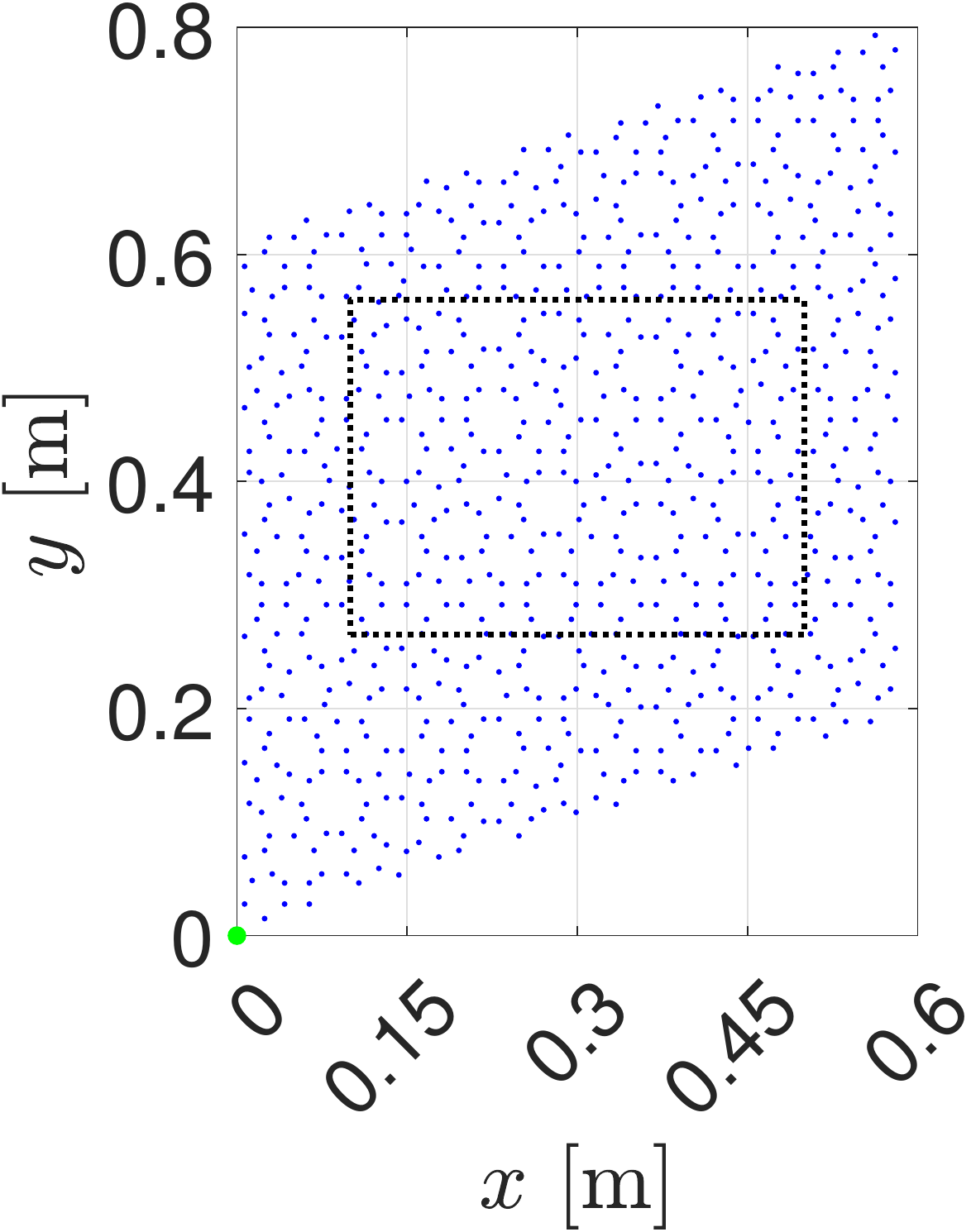}\hfill	
	\includegraphics[
	trim = 7cm 0cm 7cm 0cm,clip,
	width=0.395\textwidth]{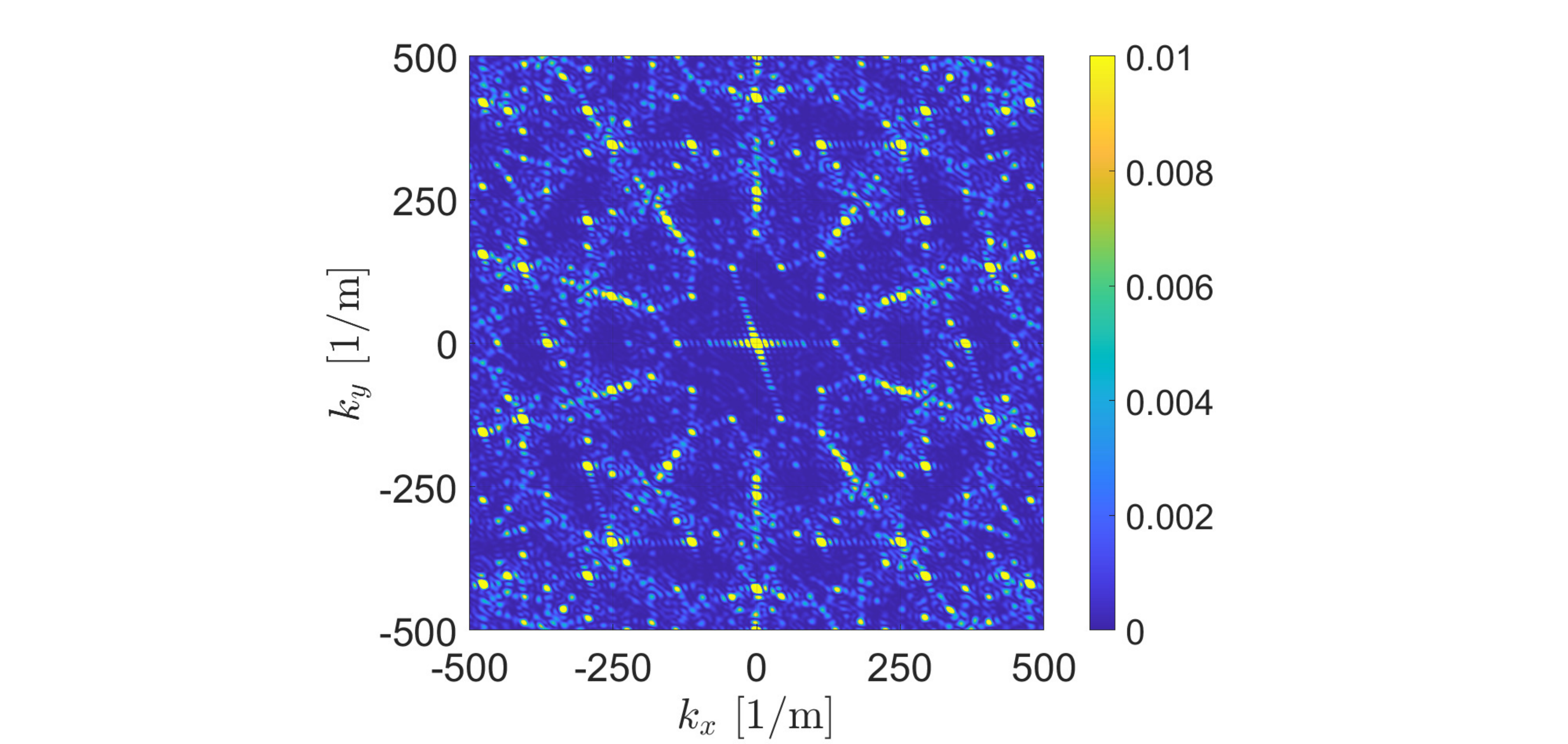}\hfill
	\vspace{1cm}
	\includegraphics[
	width=0.32\textwidth]{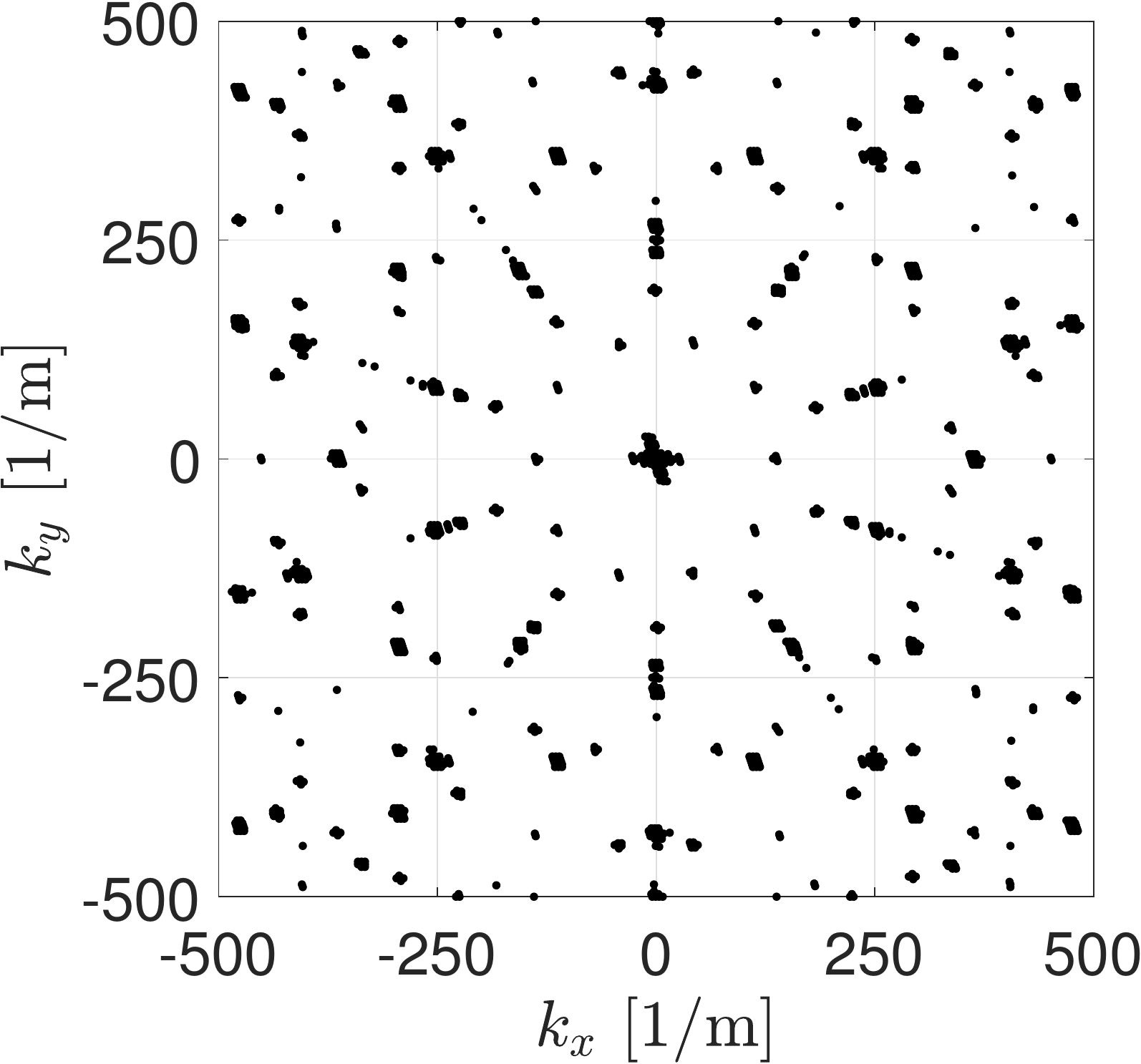}
	\vspace{0cm}
	\vfill
	\includegraphics[
	trim = 0cm 0cm -1cm 0cm,clip,  
	width=0.25\textwidth]{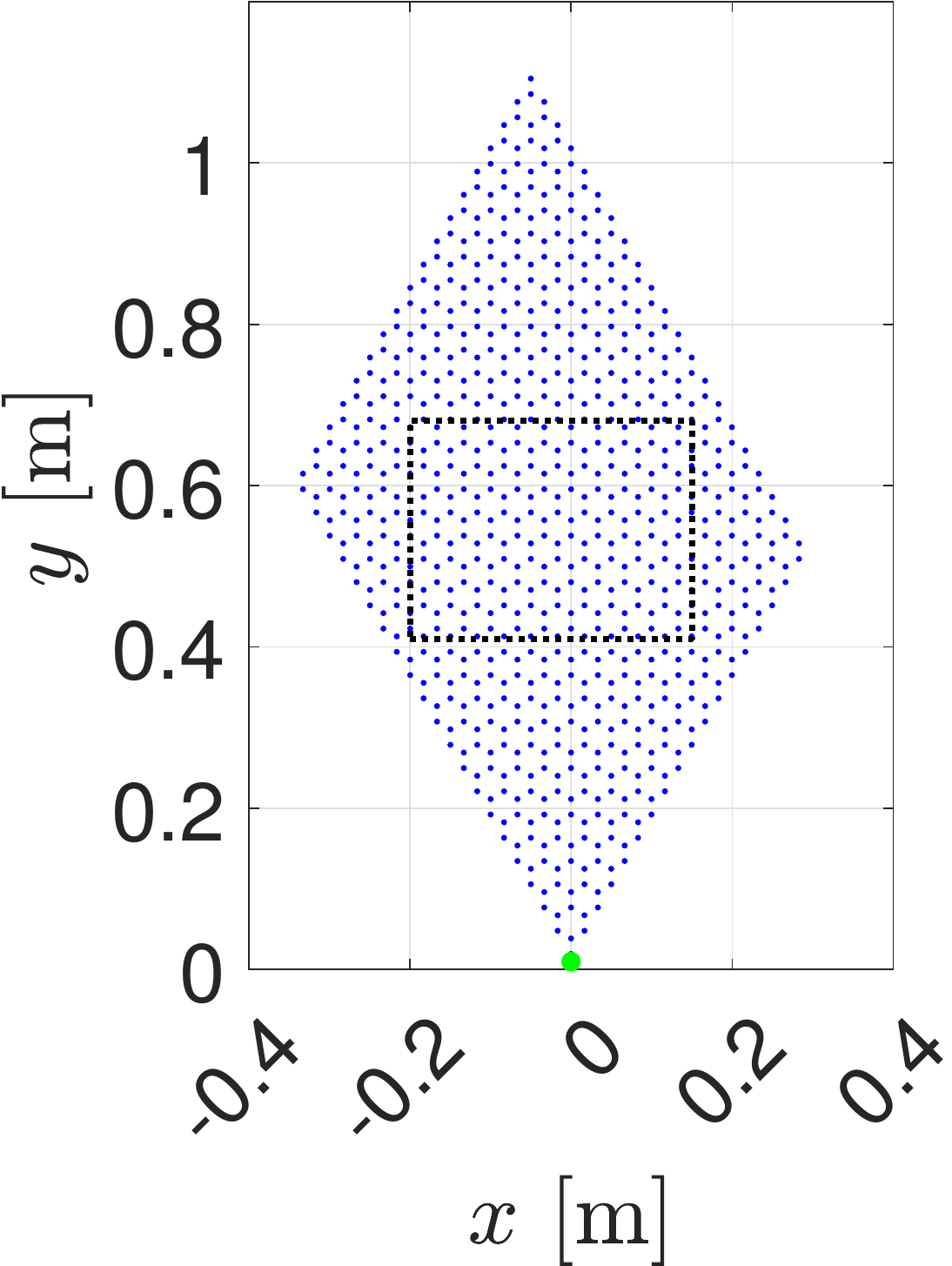}\hfill	
	\includegraphics[
	trim = 7cm 0cm 7cm 0cm,clip,
	width=0.395\textwidth]{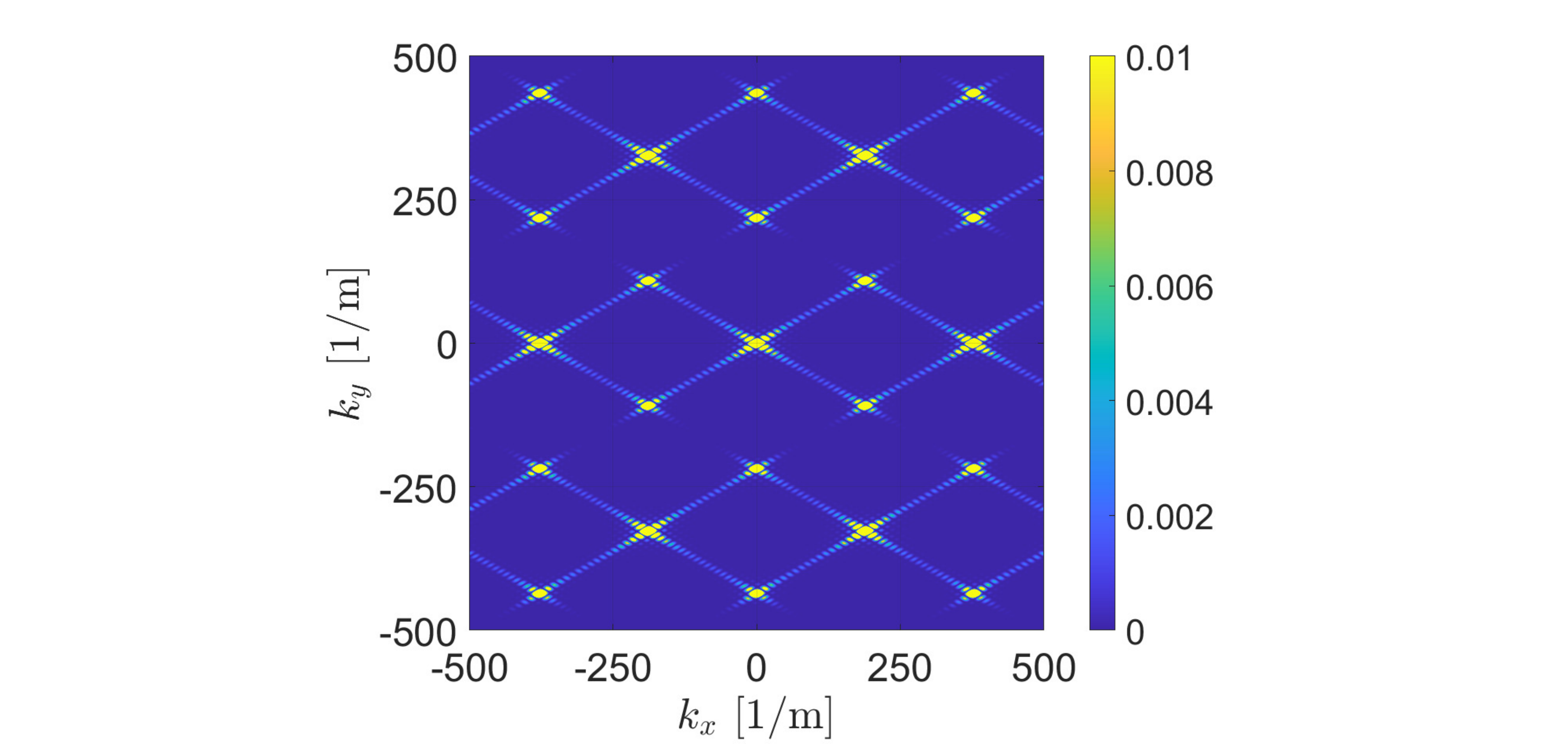}\hfill
	\includegraphics[
	width=0.32\textwidth]{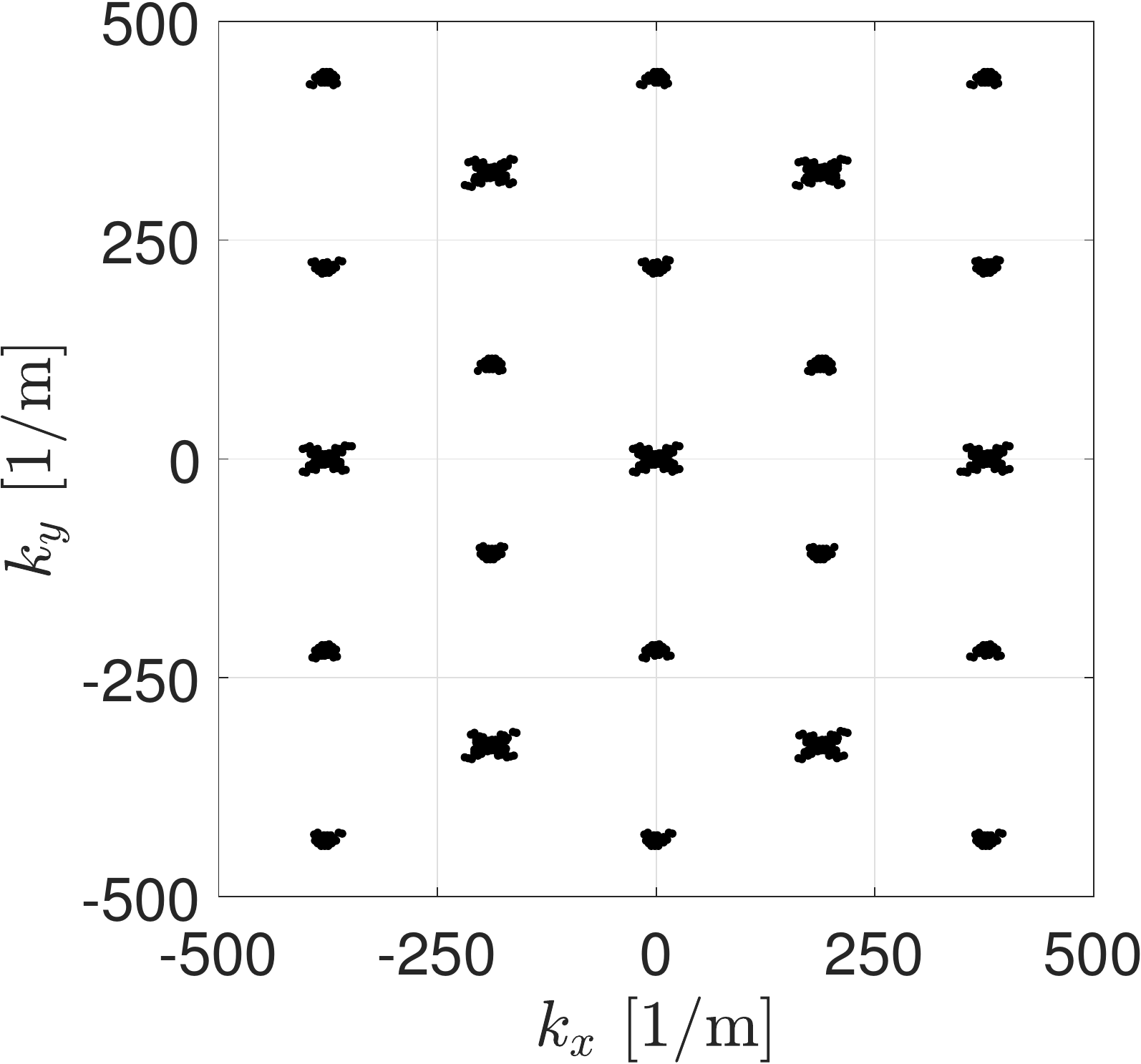}
	\begin{picture}(0,0)(0,0)
	\put(-410,170) {$(a)$}
	\put(-250,170) {$(b)$}
	\put(-75,170) {$(c)$}
	\put(-410,-10) {$(d)$}
	\put(-75,-10) {$(f)$}
	\put(-250,-10) {$(e)$}
	\end{picture}
	\caption{\label{fig:distro} (a): Distribution of Penrose-like scatterers  obtained by decomposing $n=6$ times a thick rhombus with side length $\ell_0=0.617$ m. (b): Fourier transform of the structure factor (Eq. \eqref{eq:form_factor}) corresponding to the  distribution of scatterers in panel (a). (c): A set of local maxima in panel (b) corresponding to Bragg peaks in reciprocal space of the P3 lattice. The maxima are obtained by selecting  wave vectors (black dots) for which the structure factor exceeds a given threshold $\alpha=0.05$.  (d): Equivalent honeycomb cluster of scatterers. Panels (e) and (f) have been obtained similarly to panel (b) and (c) for the  arrangement in panel (d). The  areas enclosed by the dashed rectangles in panels (a) and (d) represent probing regions of the structured plates.}
\end{figure}

\subsection{Distribution of scatterers}
By using the recurrence relation \eqref{eq:recurrence} with seeds $N^{(0)}_{\rm A}=0$ and $N^{(0)}_{\rm B}=2$, it is possible to verify that the total number of scatterers associated with a decomposition level $n=6$ is $N_s=754$ \footnote{Eq. \ref{eq:recurrence} can be recast into the  matrix form  ${\bf N}^{(n)}={\hat{M}}^{n-1} {\bf N}^{(1)} $, with 
	\begin{equation} 
	\hat{M} = 
	\begin{pmatrix}
	1 & 1\\
	1 & 2
	\end{pmatrix}.
	\end{equation} }. The resulting spatial distribution of scatterers in the truncated P3 Penrose lattice  is represented in Fig.  \ref{fig:geom}(a) by the blue dots.

The average scatterers' density $N_s/S$, with $S$ being the surface of the plate, can be used  to define a periodic counterpart of the proposed aperiodic plate.  For illustrative purposes, we opt for a graphene-like lattice because of the two-scatterers per unit cell pattern reminiscent of our P3 Penrose arrangement . In addition, the underlying triangular lattice features isotropic low-frequency dispersion compared to other Bravais lattices in two-dimensions \cite{tallarico2017tilted}. The resulting graphene-like structure has been given analysed for the first time by Torrent {\rm et al.} \cite{torrent2013elastic}, with a focus on the emergence of Dirac-like dispersion and associated dynamic response of waveguides.  

In order to provide a meaningful comparison, especially in view of the  finite plate response, we aim at preserving the total mass of the studded plate as well as the average density of scatterers.  
Approximately the same number of scatterers is given by the choice $N_1,~N_2\approx N_s/2$, where $N_1,N_2\in \mathbb{N}$  represent the number of unit cells along the directions identified by the primitive vectors of the triangular lattice. Therefore, we guarantee the same total mass by requiring that the triangular lattice primitive vectors have magnitude $\bar{\ell_0}=(2 S/(\sqrt{3}N_1 N_2))^{1/2}$. Focusing on the specific case represented in Fig. \ref{fig:distro}, $N_1 = 18$ and 
$N_2 = 21$ represents a suitable choice, corresponding to primitive lattice vectors'  magnitude $\bar{\ell_0}=0.033~{\rm m}$. By choosing the primitive vectors to be ${\bf t}_{j}=\bar{\ell_0}/2((-1)^{j-1},\sqrt{3})^{\rm T}$, with $j=1,2$, the reciprocal lattice vectors are ${\bf G}_j=2\pi/(\bar{\ell_0} \sqrt{3}) (\sqrt{3}(-1)^{j-1},1 )^{\rm T}$. 

\subsection{Structure factor and diffraction patterns \label{sec:structure_factor}}
In reciprocal space, the structure factor is defined as the modulus squared of the Fourier transform of the scattering potential (see the right hand side of Eq. \ref{eq:governing-eq}), normalized by the amplitude $f(\beta)$, \emph{i.e.}
\begin{equation}\label{eq:form_factor}
{\cal F}({\bf k}) = \left|\frac{1}{N_s}\sum_{j=1}^{N_s}\exp(i {\bf k}\cdot {\bf r}_j)\right|^2, 
\end{equation}
where the index $j=1,\cdots,N_s$ runs over the Penrose arrangement of coordinates ${\bf r}_j$ in the plane, and ${\bf k}=\left(k_x,k_y\right)^{\rm T}$, is the wave-vector. Such quantity is customarily used in condensed matter physics to analyse particle or wave diffraction experiments, as it is proportional to the scattering cross-section. As the name suggests, the structure factor contains important crystallographic information on the atomic distribution. Its evaluation allows to the diffraction intensity in reciprocal space, thus possibly allowing an approximate identification of the Bragg peaks, induced by a finite but large distribution  of scatterers. Bragg peaks are special wave-vectors  at which strong dispersive behaviour, such as the opening of band-gaps, is expected. The evaluation of structure factors is customarily used in the research community working on aperiodic media \cite{braake2013aperiodic}.       
\color{black}	
\subsection{Frequency estimation of band gaps}
At low frequency, the effect of the scatterers can be accounted for by introducing an effective material density 
\begin{equation}\label{eq:homo-density}
	\rho = \rho_s + \frac{m N_s}{S h},
\end{equation} 
where $m$ is the mass of each scatterer and $S$ is the surface of the enclosing plate. The low-frequency dispersion for flexural waves in the long-wavelength regime can be easily obtained by evaluating the dispersion relation in Eq. \eqref{eq:equation}, using Eq. \eqref{eq:homo-density} as mass density.
Using Eq. \eqref{eq:equation}, one can estimate  frequencies for flexural waves corresponding to  special points in reciprocal space as
\begin{equation}\label{eq:BG}
\nu^{*} \approx \frac{1}{2\pi} \sqrt{\frac{D h}{\rho}} \left|{\bf K}\right|^2,
\end{equation} 
where the density $\rho$ has been introduced in Eq. \eqref{eq:homo-density}. We underline that Eq. \eqref{eq:BG} is not the exact expression for  Bloch frequency at a specific point in the reciprocal space for the periodic structure.  As shown in Sec. \ref{sec:periodic}, although Eq. \eqref{eq:BG} is asymptotically equivalent to the low-frequency dispersion behaviour of the periodic counterpart, as the wave frequency increases dispersive effects stemming from periodicity inevitably lead to pronounced departures from the dispersion of a homogeneous plate. Nevertheless, estimations based on \eqref{eq:BG} will help us pin-point  comparative effects in the dynamic responses of the P3 Penrose cluster and the quasi-statically equivalent honeycomb one.


\section{Results and discussion \label{sec:results}} 
\subsection{Effect of the scatterers finite-size on the dispersion of periodic counterparts \label{sec:periodic}} In order to single out  the scatterers' finite size effects, we opt to follow a FE approach, due to the ability of the method to reliably cope with coupling of the elastic field within the plate with the rotational and translational degrees of freedom of the scatterers (see Sec. \ref{sec:fe_formulation}).  

\begin{figure}[h!]
	\centering
	\vspace{12cm}
	\begin{picture}(0,0)(0,0)
	\put(-250,50) {\includegraphics[trim=0cm 0cm 0cm 0cm, clip,width=0.4\textwidth]{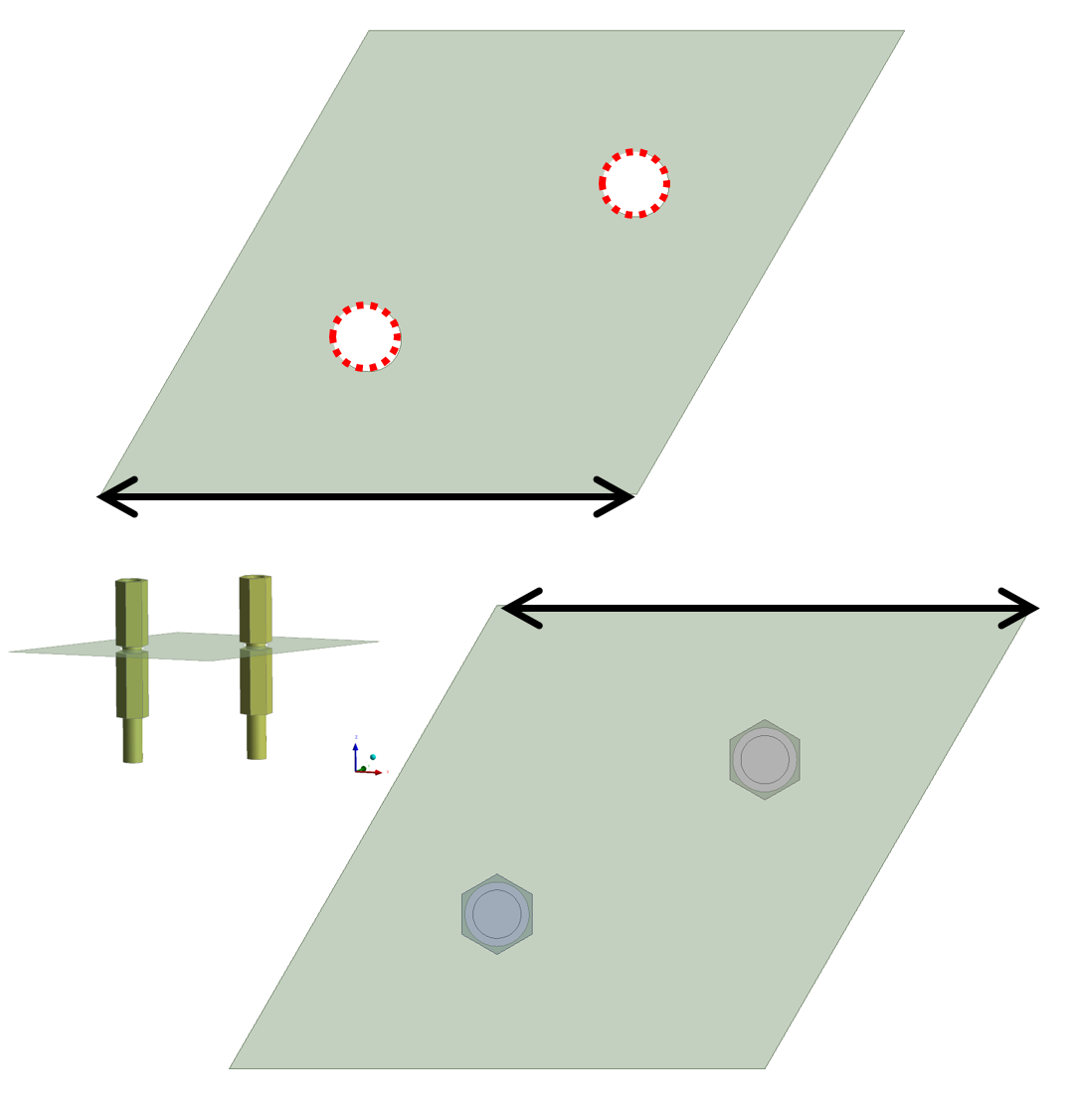}}
	\put(-50,0) {\includegraphics[trim=0cm 0cm 0cm 0cm, clip,width=0.33\textwidth]{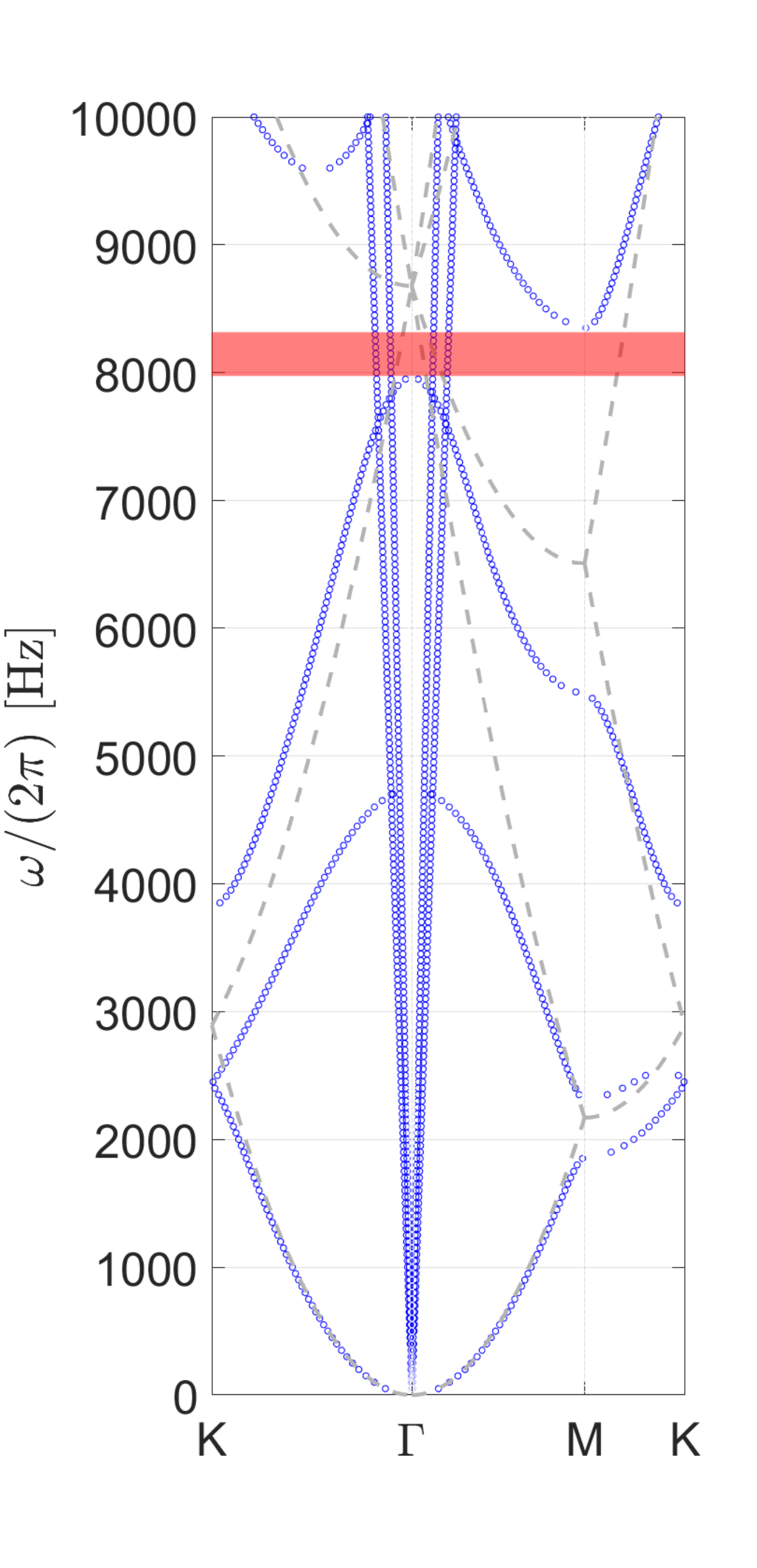}}
	\put(100,0) {\includegraphics[trim=0cm 0cm 0cm 0cm, clip,width=0.33\textwidth]{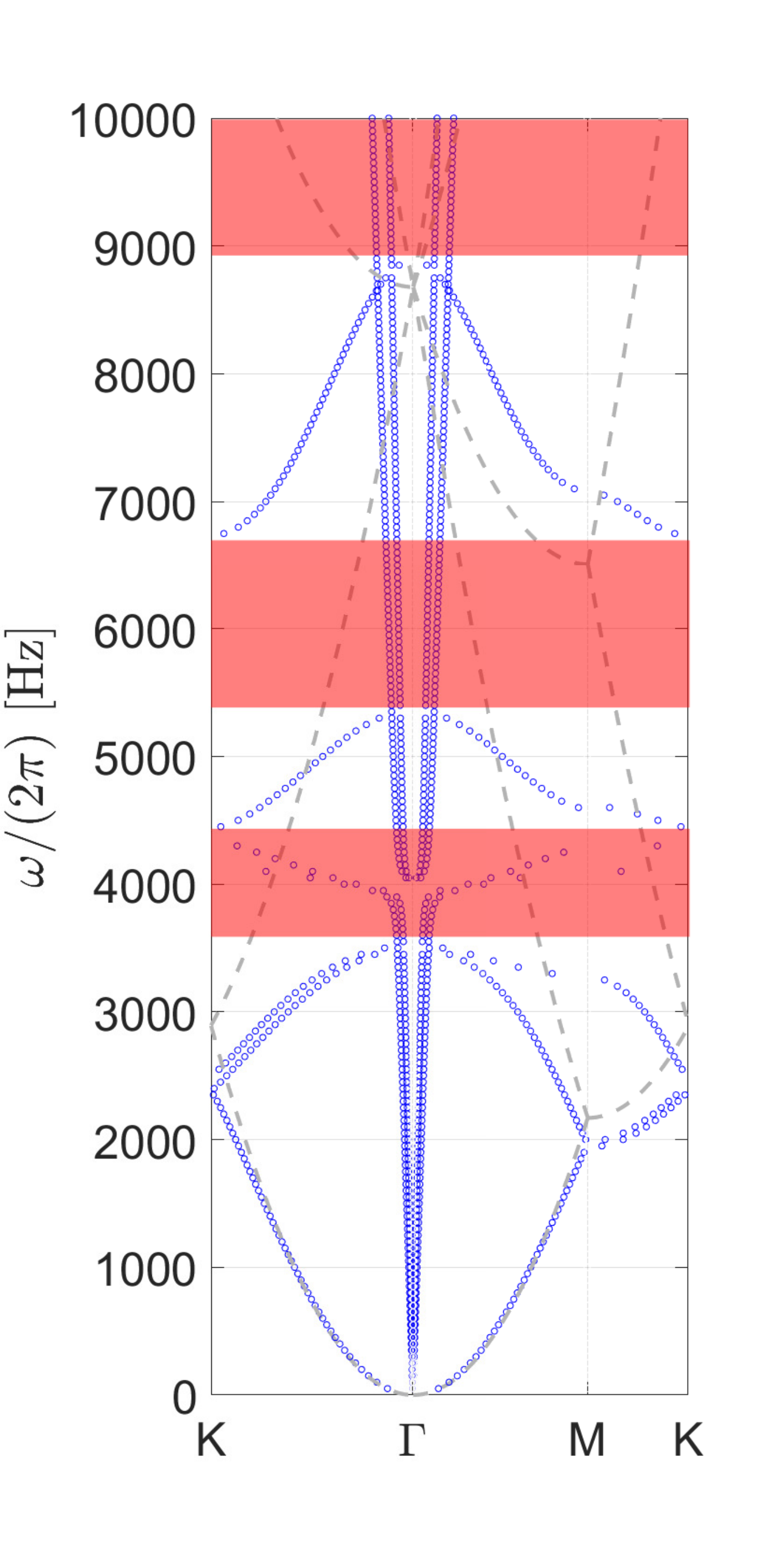}}
	\put(-120,180){UC1}
	\put(-200,50) {UC2}
	\put(-140,145){$\bar{\ell}_0$}
	\put(-150,0) {${\bm (a)}$}
	\put(20,0) {${\bm (b)}$}
	\put(170,0) {${\bm (c)}$}
	\end{picture}
	\caption{\label{fig:disp}(a): Two honeycomb unit cells (UC1 and UC2). UC1 consists of two drilled holes with edge mass distribution whereas in UC2 the holes host two rigid studs of mass $m$ (see inset). The  radius of the holes is $r= 1.5$ mm and the mass of the scatterers is $m = 3.2$ g. Panel (b) and (c) show the dispersion diagrams of UC1 and UC2, respectively, projected over the irreducible Brillouin zone (IBZ) in Fig. \ref{fig:mobility_penrose_ana}(a).   The blue  circles represent the (real) Bloch-Floquet wave-vector along the boundaries of the IBZ. The grey dashed lines represent the dispersion of a homogeneous plate with density as in Eq. \eqref{eq:homo-density}, folded in the first IBZ of the triangular lattice.}
\end{figure}
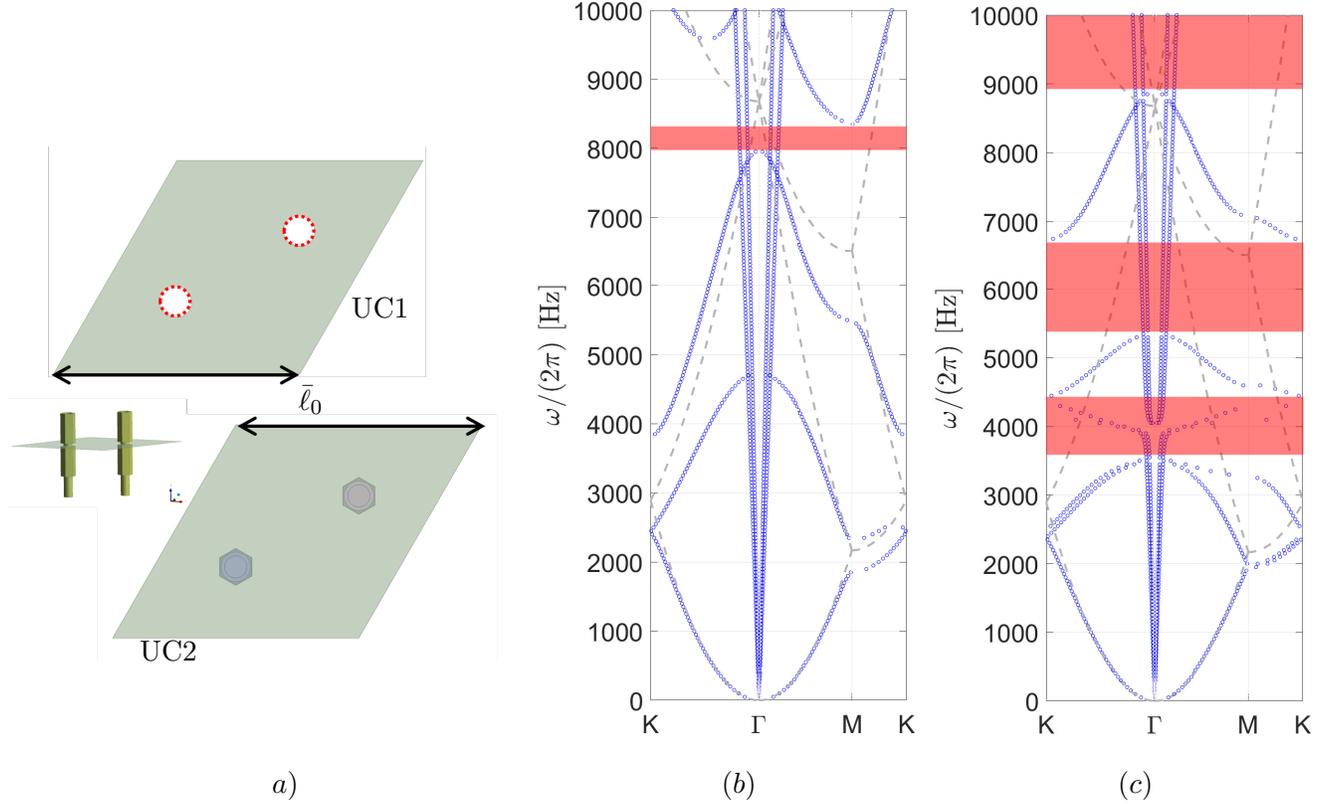

In the present section, we focus on the dispersive properties of two unit cells in Fig. \ref{fig:disp}(a) (UC1 and UC2) whose repetition leads to the honeycomb lattice.  
In UC1 we consider an edge distribution of mass (3.2 g in total) along the circumference of each hole (see red dashed circles) with radius 1.5 mm. In UC2, the scatterers are modelled as rigid studs which pierce the plate through the same holes as in UC1. The precise geometry and materials parameters can be accessed on the website of the manufacturer \cite{spacer}. In table \ref{tab:studs}, we report the key physical parameters of the studs and of the plate considered in the finite-element formulation. Specifically, we introduce the key stud paramaters: the mass ($m$), the moments of inertia around  the $x-$, $y-$ and $z-$axes ($I_x$, $I_y$ and $I_z$, respectively), and the vertical offset of the centre of mass from the mid-surface of the plate ($h_z$). This latter parameter models the fact that the stud is not symmetric about the midplane of the plate. In addition, the unit cell length $\bar{\ell}_0$ has been derived from the considerations in Sec. \ref{sec:structure_factor}. 
\begin{table}
	\begin{tabular}{ccccccccccc}
		\toprule
		\multicolumn{5}{c}{Stud} &	\multicolumn{1}{c}{} & \multicolumn{4}{c}{Plate}\\
		\cmidrule{1-5}  \cmidrule{7-10} \\
		$m$ [g] & $I_x$ [kg $\rm m^2$] & $I_y$ [kg $\rm m^2$] & $I_z$ [kg $\rm m^2$] & $h_z$ [mm] & & $E_s$ [GPa] & $\nu_s$ [-] & $\rho$ [kg/$\rm m^3$] & $h$ [mm] \\
		\midrule
		3.2 & $1.63 \cdot 10^{-7}$ & $1.63 \cdot 10^{-7}$ & $8.13 \cdot 10^{-9}$& -3.0  & & 200&0.3&7850&1.014 \\
		\bottomrule
	\end{tabular}
	\caption{\label{tab:studs} Physical parameters of the rigid studs and of the homogeneous plate.}
\end{table}
The dispersion diagrams considered in this section have been calculated using a direct Bloch-Floquet technique \cite{mace2008modelling,tallarico2020superelement}.
The comparison of the two dispersion diagrams in Figs \ref{fig:disp}(b) and \ref{fig:disp}(c) shows that, in the long wavelength regime (i.e. below 1000 Hz) the dispersion diagram is well captured by that of a homogeneous plate with density as in Eq. \eqref{eq:homo-density}, which have been folded onto the boundaries of the IBZ (see grey dashed lines) for ease of comparison. 

Unsurprisingly, as the frequency increases, marked departures from the homogeneous dispersive behaviour arise. Firstly, the dispersion of UC1 does not feature any coupling between in-plane and flexural waves. The absence of coupling can be inferred by the purely linear dispersion branches in Fig. \ref{fig:disp}(b) associated with plane-stress modes. On the other hand the dispersive properties of the UC2 feature  hybridisation of in-plane and flexural dispersion branches around 4000 Hz, a phenomenon stemming from the activation of the rotational inertias and from the asymmetry of the studs with respect to the midplane. This leads to an effective band-gap for flexural waves highlighted by the corresponding red rectangle and further illustrated in Sec. \ref{sec:experiment}.

In addition, UC1 features a Bragg-scattering band gap around $\nu^*({ |\bf G}_1|)$, as highlighted by the red-shaded rectangle in Fig. \ref{fig:disp}(b). By contrast, Fig. \ref{fig:disp}(c) shows more pronounced band-gaps for flexural waves (see the width of the red-shaded rectangles). The higher band gap in frequency is clearly controlled by the mass of the scatterers,  whereas the lower ones are controlled by the rotational inertias  coupling with the plate rotation fields (i.e. with the derivatives of the flexural field $\psi({\bf r})$).      
 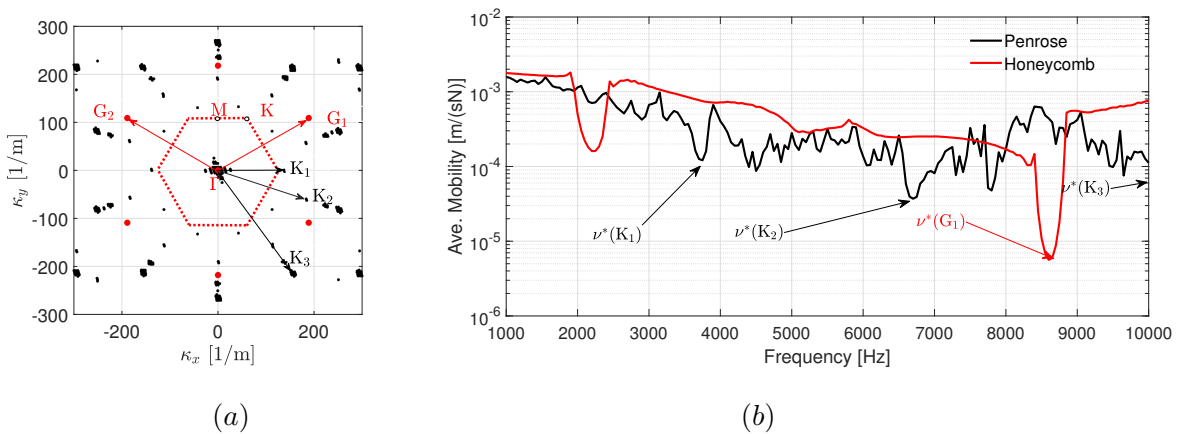
\begin{figure}[h!]
	\centering
	\includegraphics[trim=10.5cm 0cm 12cm 0cm, clip,width=0.33\textwidth]{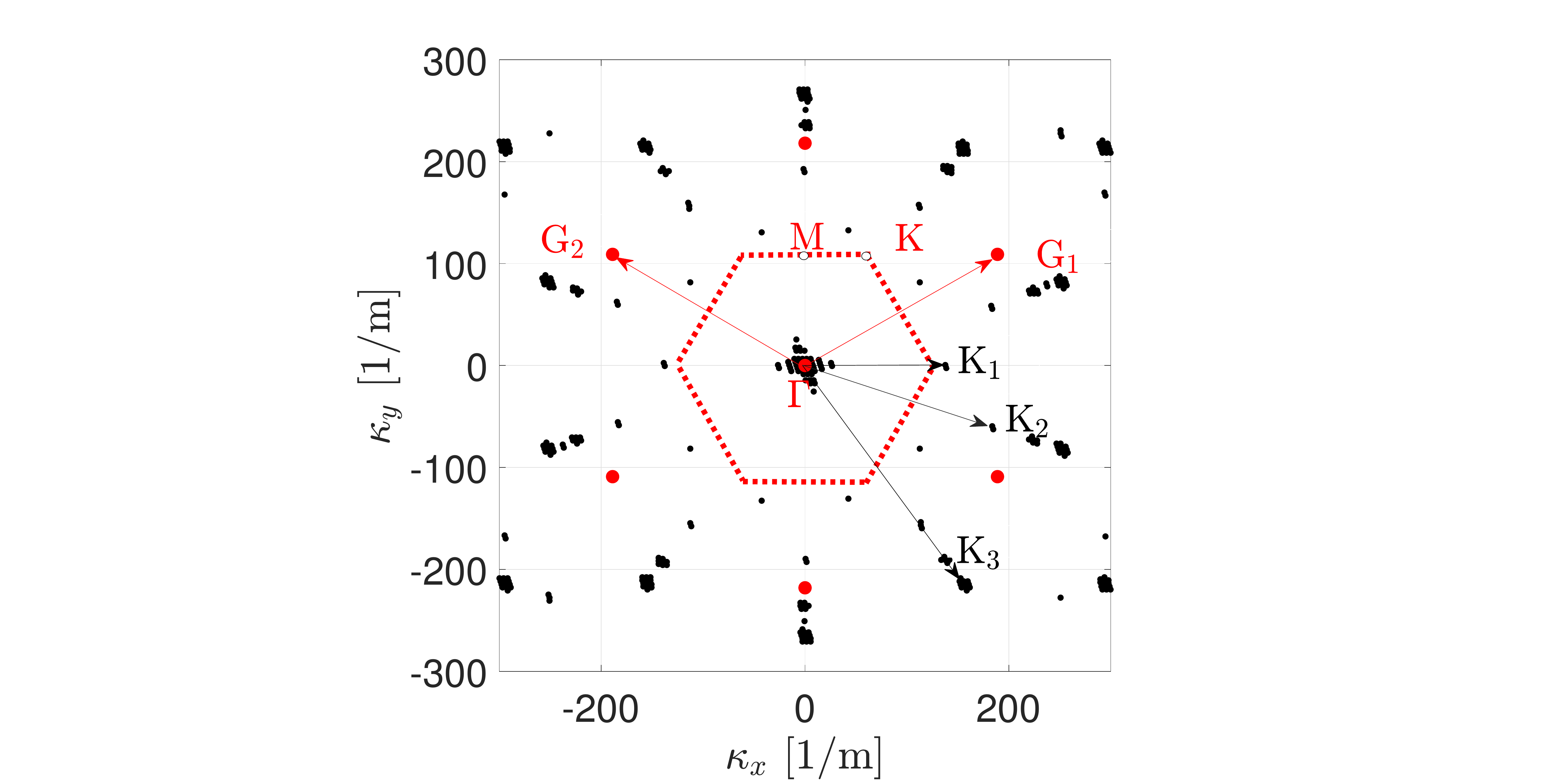}\hfill
	\includegraphics[
	width=0.67\textwidth]{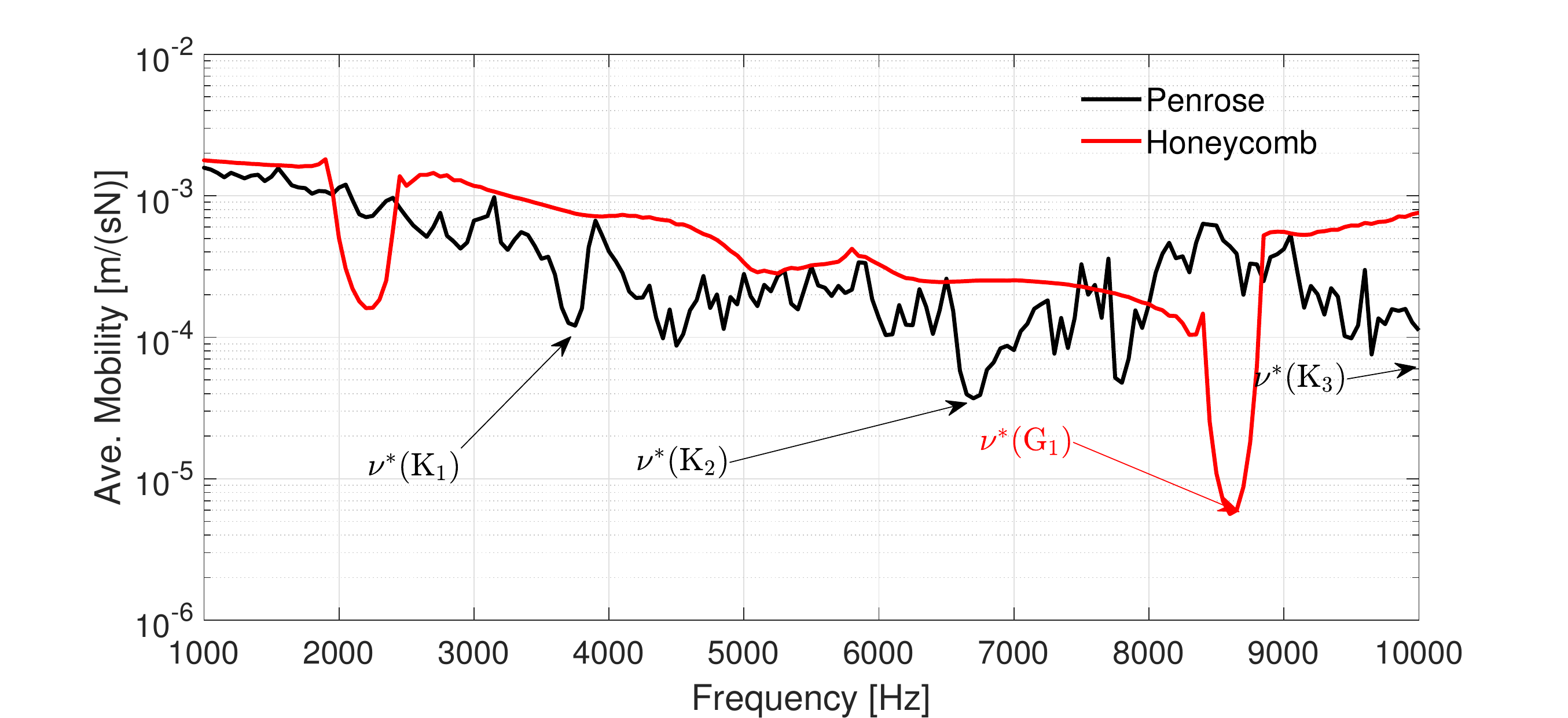}
	\begin{picture}(0,0)(0,0)
	\put(-150,0) {$(a)$}
	\put(50,0) {$(b)$}
	\end{picture}
	\caption{\label{fig:mobility_penrose_ana}Panel (a) shows the approximate Bragg peaks of the finite Penrose point-masses cluster (black dots) and the Bragg peaks of the honeycomb point-mass cluster (red lines) with equivalent density. The dashed hexagon is the edge of the first Brillouin zone of the triangular lattice, where the high-symmetry points M, K, and ${\rm \Gamma}$ are also marked. Panel (b) shows the clusters' mobility to a point force located at the origin of the coordinate system. Black and red lines refer to the Penrose and the honeycomb cluster, respectively. The frequencies corresponding to the respective Bragg peaks (see panel (a)) are also shown.}
\end{figure} 
\subsection{Bragg scattering by phononic clusters within an infinite thin plate}
In this section, we study the dynamic response of the clusters of scatterers represented in Figs \ref{fig:distro}(a) and \ref{fig:distro}(d). 
By solving Eq. \eqref{eq:governing-eq} with the method outlined in \color{black} Appendix \ref{sec:appendix}\color{black}, it is possible to probe a finite cluster of point masses in a infinite thin plate to an external incident field. We deliberately focus on  a cluster made of point masses without resonant effects, for which \cite{torrent2013elastic,wu2008evidence}
\begin{equation}
 f_i(\beta)= \frac{m}{\rho h} \beta^4,~~~~\forall i=\{1,\cdots,N_s\}, 
\end{equation}  
in Eq. \eqref{eq:governing-eq}, with $m$ being the mass of the scatterers. The dispersion diagram associated with the periodic honeycomb cluster is in Fig. \ref{fig:disp}(b).  For simplicity, we assume that the incident field originates from a point source generated by a unit force, although other solutions compatible with the Kirchhoff plate can be chosen. The cluster is arranged as in Fig. \ref{fig:distro}(a) and the point source is located at the origin of the coordinate system (see green circle therein).

 A measure of the dynamic response, directly accessible in vibrometric experiments, is the average mobility over a probing surface, that is 
\begin{equation}\label{eq:ave_mob}
{\cal M}(\omega) =\frac{\omega<|\psi({\bf r},\omega)|>}{F_0(\omega)},
\end{equation}
where $F_0(\omega)$ is the time-Fourier transform of the input force on the plate, $\psi({\bf r},\omega)$ is time-harmonic flexural field,  $<\cdot>=1/S\int {\rm d}^2{\bf r}(\cdot)$ is the average operator over a surface of area $S$, and $|(\cdot)|$ denotes the absolute value of a complex number. In computations, we assume $F_0(\omega)=1$. The average mobility within a probing region (dashed rectangle in Fig. \ref{fig:distro}(a)) of the Penrose cluster is represented in Fig. \ref{fig:mobility_penrose_ana}(b) by the black solid curve, whereas the red line represents the average mobility of a honeycomb cluster region (dashed rectangle in Fig. \ref{fig:distro}(d)).  

The two phononic crystals feature strong Bragg peaks, represented in Fig. \ref{fig:mobility_penrose_ana}(a).  The black points refer to the Penrose lattice and are identical to those in Fig. \ref{fig:distro}(c), whereas the red dots are the Bravais nodal points in reciprocal space of the triangular lattice. The scattering of flexural waves   within the Penrose cluster results in a plethora of local minima in mobility, whose frequencies are lower than $\nu^{*}(|{ G}_1|)$. Pronounced minima are observed at the special frequencies $\nu^{*}(|{\rm K}_1|)$,$\nu^{*}(|{\rm K}_2|)$ and $\nu^{*}(|{ K}_3|)$ for the Penrose cluster, identified in section \ref{sec:structure_factor}.  This results in a overall pronounced attenuation of up to 1 order of magnitude in the Penrose cluster  compared to the honeycomb cluster up to 8000 Hz  (see Fig. \ref{fig:mobility_penrose_ana}(b)).  It is worthwhile noting that the mobility of the honeycomb cluster features attenuation consistent with the band-gap that has been predicted within the periodic counterpart (\emph{cf.} Figs \ref{fig:disp}(b) and \ref{fig:mobility_penrose_ana}(b)). The dip in mobility around 2000 Hz for the honeycomb cluster deserves a special mention. As shown in the dispersion diagram in Fig. \ref{fig:disp}(b), the lattice features a partial band-gap at M, \emph{i.e.} for waves propagating along $y$. This partial band-gap reflects flexural direct waves coming from the source, while allowing indirect waves (i.e. scattered from the edge of the cluster) to reach the computational domain, leading to a steady-state dip in mobility of lower efficiency compared to that resulting from the total band-gap.  
\subsection{Penrose studded plate mobility and localization \label{sec:experiment}}

 \begin{figure}[h!]
	\centering
	\includegraphics[width=1\textwidth]{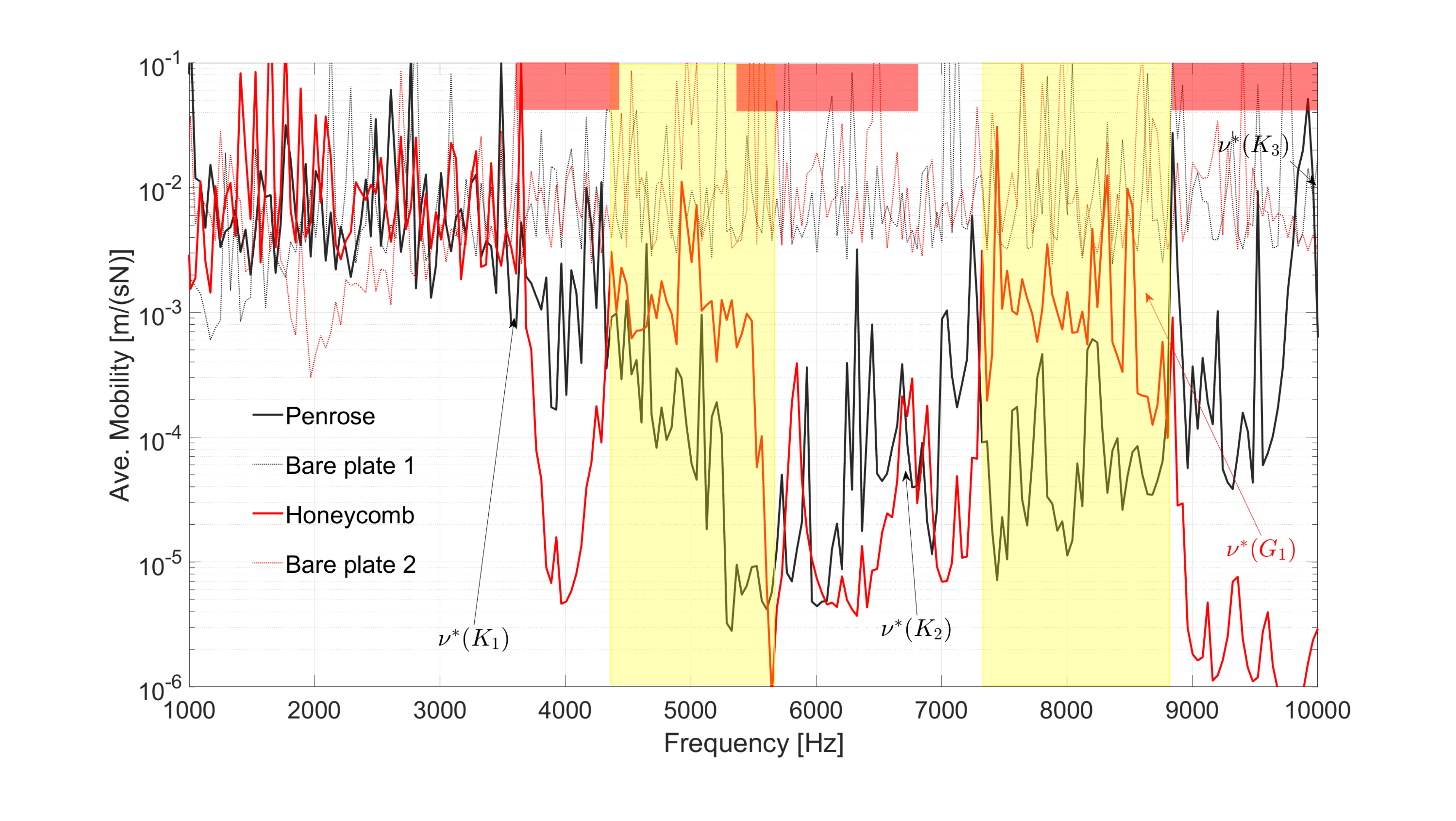}
	\caption{\label{fig:mobility-psp-hsp}  FE average mobility  of the two finite plates with traction-free boundary conditions. The black and red solid lines refer to the Penrose and honeycomb studded lattices, respectively.  The material parameters used for both plates are reported in Tab. \ref{tab:studs}. The red and black dotted lines represent the average mobility of homogenised plates (honeycomb and Penrose, respectively), with homogenous mass density as in Eq. \eqref{eq:homo-density}.}
\end{figure} 
For a Penrose studded plate, a unit cell in the Bloch-Floquet sense does not exist.  On the other hand, insights can be gained by analysing the dynamic response of a finite plate, subjected to a harmonic loading and to physical boundary conditions at the edges of the sample. The most straightforward way to model the aforementioned situation is by using the FE method. Specifically, we use a time-harmonic formulation with unit forcing vector oriented normal to the plate. We repeat the computation for 200 equally-spaced frequencies between 1k Hz and 10k Hz, which being independent to one another can be readily parallelised for improved computational time efficiency. The rigid scatterers assumption for the studs is justified by an independent modal computation carried over a single elastic stud made out of Brass and showing that its first non-trivial eigenmode is above 20 kHz, therefore far away from our frequency range of interest. The contact between the rigid-body scatterer and the plate is of the bonded type, which implies continuity of rotations and displacement at the plate/scatterers junctions.  
    
In the same spirit as in Fig. \ref{fig:mobility_penrose_ana}, Fig. \ref{fig:mobility-psp-hsp} compares the average mobility of the two finite plates subjected to a unit point force and complemented with traction-free boundary conditions at the boundaries of the samples. The black and red curves refer to the Penrose and honeycomb studded plates, respectively. Each stud is modelled as described in the context of Fig. \ref{fig:disp}(c). The red rectangles show the band-gaps within the periodic plate, as shown in Fig. \ref{fig:disp}(c), which is consistent with the corresponding low average mobility in those regions (see red solid curve). As highlighted by the yellow-shaded areas in Fig. \ref{fig:mobility-psp-hsp}, the mobility of the Penrose plate is again lower than the periodic counterpart at lower frequencies compared to $\nu^{*}(|{\rm G}_1|)$. In order to single out the effect of periodic and aperiodic order, we report the average mobility of homogeneous finite plates (Bare plate 1 and Bare plate 2) with elastic parameters as in Tab. \ref{tab:studs} and mass density as in Eq. \eqref{eq:homo-density}. Such homogenous counterparts only differ by the overall shape of the plate. We observe that, despite the bare plates differ by the overall shape, this affects only the fine modal behaviour but not the mobility order of magnitude in such a broad frequency range. By direct comparison of the solid and dotted curves in Fig. \ref{fig:mobility-psp-hsp}, the studded plates result in strong attenuation compared to the homogenous counterparts. Moreover, in the frequency regime from 3000 Hz up to 9000 Hz, the realisation of the Penrose plate analysed in the current paper results in a consistently lower mobility. In turns, this results in the presence of Bragg peaks associated with aperiodic order.           

In order to test the predictive power of the numerical models, we have commissioned a plate with holes as per the blue dots in Fig. \ref{fig:distro}(a). We have subsequently screwed the studs described in Sec. \ref{sec:periodic} into the holes, making sure that they are tight thus mimicking bonded conditions. 
 The resulting studded plate has been subsequently analysed via laser Doppler vibrometry. Specifically, we drove a shaker using a frequency sweep (1k-8k Hz). The resulting force between the stinger of the shaker and the plate, applied to the origin of the coordinate system in Fig. \ref{fig:distro}(a), is measured by interposing a force sensor. The resulting force signal is used as reference for the laser Doppler vibrometer thus allowing a direct measurement of the mobility over a scanning region.   

\begin{figure}[h!]
	\includegraphics[width=0.42\textwidth]{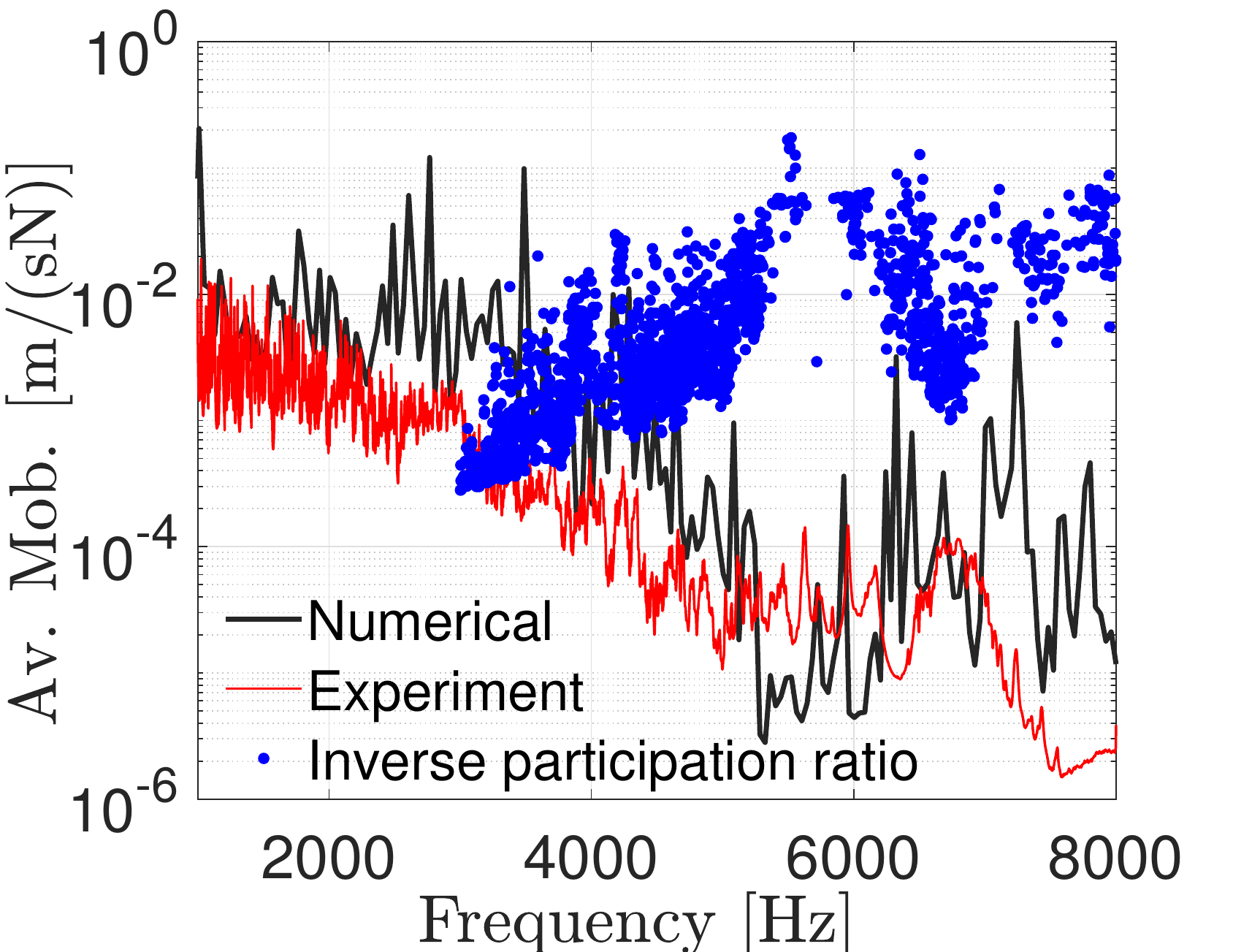}
	\hfill
	 \includegraphics[width=0.57\textwidth]{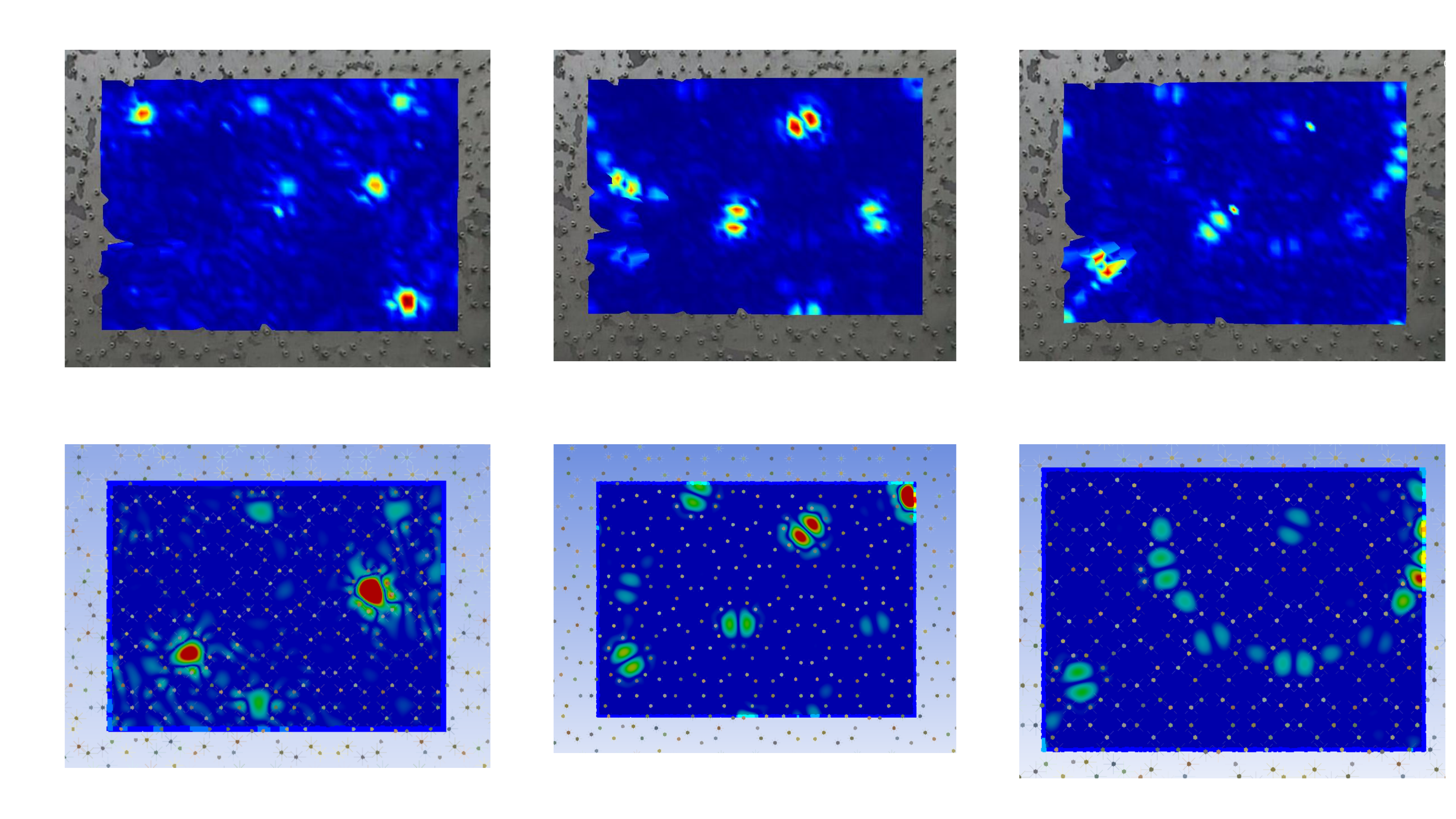}
	$~~$
	\begin{picture}(0,0)(0,0)
	\put(-140,10) {${\bm (a)}$}
	\put(-20,94.5) {${\bm (b): 5623}$ Hz}
	\put(70,95) {${\bm (c): 7297}$ Hz}
	\put(155,95) {${\bm (d): 7462}$ Hz}
	\put(-20,20) {${\bm (e): 5467}$ Hz}
	\put( 70,20) {${\bm (f): 7180}$ Hz}
	\put(155,20) {${\bm (g): 7339}$ Hz}
	\end{picture}
	\caption{\label{fig:mobility-exp-num } Comparison between the numerical average mobility (black solid line) and  the average mobility obtained from vibrometric experiments (red solid line).  The blue dots in panel (a) show the inverse participation ratio for the modal response of the Penrose plate (see Eq. \eqref{eq:IPR}).   Panels (b)-(d) show time-Fourier transform of the experimentally observed localised modes   whereas panels (e)-(g) show the corresponding FE time-harmonic predictions.}
\end{figure}
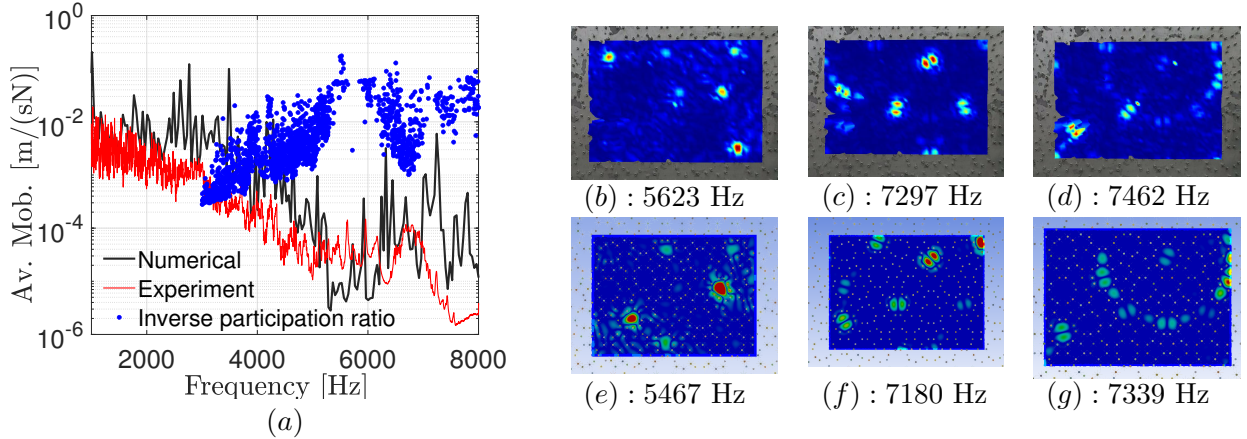

In Fig. \ref{fig:mobility-exp-num }(a), we report a comparison of the experimental and numerical mobility in similar regions of the Penrose studded plate (red and black solid curves, respectively). The comparison shows a fair agreement of numerical and experimental results. The space-resolved experimental (Figs. \ref{fig:mobility-exp-num }(b)-\ref{fig:mobility-exp-num }(d)) and numerical (Figs. \ref{fig:mobility-exp-num }(d)-\ref{fig:mobility-exp-num }(f))  waveforms compare very well. Panels (b)-(f) show highly localised far away from the source of vibration resulting in very attenuating response of the Penrose plate. 

In order to gain further insights on the effect of Bragg-scattering within the Penrose plate, we have extracted the inverse participation ratio over the flexural components of the structural modes of the plates. By solving a modal analysis in ANSYS, and extracting the associated modes in the frequency range from 3000 Hz to 8000 Hz, the inverse participation ratio (IPR) for the mode with frequency $\nu_i$ is defined as \cite{maximo2015spatial}
\begin{equation}\label{eq:IPR}
\Pi(\nu_i) = \sum_{n=1}^{N} \Psi_n^4(\nu_i),
\end{equation}  
where $\Psi_n$, $n=\{1,\cdots,N\}$ denote the nodal flexural mode of the plate and $N$ the total number of nodes. By assigning the mode normalization  $\sum_{n=1}^{N} \Psi_n^2(\nu_i)=1$, it is easy to realise that  $\Pi(\nu_i)$ approaches 1 for very localised modes (\emph{i.e.} only one node contributes 1, the remaining being zero) whereas $\Pi(\nu_i)\rightarrow1/N$ for delocalised modes (each node equally contributes $1/\sqrt{N}$). In our numerical experiment, $N=11085$ in the probing region. The inverse participation ratios for the modal response of the Penrose studded plate is represented by the blue dots in Fig. \ref{fig:mobility-exp-num }(a), and can be directly compared to the mobility curves. We highlight that low values of the mobility are associated with localised modes (i.e. modes for which their IPR approaches 1) as also confirmed by the space-resolved experimental observations. The existence of such localised modes is radically different from localization in periodic counterparts, associated with band-gap formation, where is observed only in the neighbourhood of the input location.   
  
  \section{Conclusions\label{sec:conclusions}}
  In conclusion, we have described analytically, numerically and experimentally the dynamic response of a finite plate comprising  scatterers arranged aperiodically according to a P3 Penrose tiling pattern (see Fig. \ref{fig:geom}). The  comparison with an equivalent (i.e. featuring similar low-frequency dispersive properties) periodic counterpart shows that the Penrose lattice is able to attenuate vibrations in a wide frequency range, well below the Bragg frequency of a honeycomb equivalent lattice. In addition, the results singled out the effect of rotational inertia whose inclusion  give rise to marked departures from the Bragg scattering of flexural wave only. 
  
  The long-range order in the P3 Penrose lattice results in strong Bragg wave-vectors in reciprocal space, which in turn influences the dynamic response of the cluster. This is evident when a non-locally resonant cluster is considered as in Fig. \ref{fig:mobility_penrose_ana}. The inclusion of rigid studs activate the rotational degree of freedom of the scatterers which is responsible for the departures from the simple picture of Bragg-scattering dominated by purely vertical forces (point mass situation). Nevertheless, similar conclusions can be drawn even for the finite plate studded with rigid inclusions. Since Bragg-type attenuation in a P3 lattice happens at lower frequency compared to the equivalent periodic counterpart, the results are relevant for structural engineering applications, where controlling the low-frequency response is critical. In addition, we have identified a wave localization frequency regime shaped by Bragg-scattering within the aperiodic realization of the Penrose P3 tessellation. The control of such Bragg-scattering induced localization in aperiodic plates may prove useful  in the context of energy harvesting or to  damp even further unwanted structural modes at desired frequencies.
\appendix
\section{Multiple-scattering theory\label{sec:appendix}}
Using the Korringa-Kohn-Rostoker (KKR) \cite{korringa1947calculation,kohn1954solution} method, given an ``incident'' field $\phi_0$ which is solution of the unperturbed problem, we can write the elastic field in an arbitrary position ${\bf r}$ within the plate as  
\begin{equation}\label{eq:kkr_solution}
\psi({\bf r})=\phi_0({\bf r}) + \int d^2{\bf r}_0~g_0({\bf r},{\bf r}_0; \beta) V({\bf r}_0)\psi({\bf r}_0),
\end{equation}
where we have introduced the Green's function in Eq. \eqref{eq:gf}, and $V({\bf r}_0)$ is the right-hand-side of Eq. \eqref{eq:governing-eq}. Specifically, the incident flexural displacement resulting from vertical load of amplitude $F_0$ at the origin of the coordinate system is 
\begin{equation}
\phi_0({\bf r})= \frac{F_0}{D}g_0({\bf r},0;\beta).  
\end{equation} 
By using $V({\bf r}_0)=\sum_{i=1}^{N_s}f(\beta)\delta({\bf r}_0-{\bf r}_i)$ in Eq. \eqref{eq:kkr_solution}, we obtain 
\begin{equation}\label{eq:kkr_solution_dpotential}
\psi({\bf r})=\phi_0({\bf r})+f(\beta)\sum_{j=1}^{N_s}  g_0(|{\bf r}-{\bf r}_j|) \psi({\bf r}_j).
\end{equation}
By taking the limit ${\bf r}\rightarrow{\bf r}_i$, the following linear system follows 
\begin{equation}
(\hat{1}-\hat{G}(\beta)){\bm \psi}={\bm \phi},
\end{equation}
where $[\hat{1}]_{ij}=\delta_{ij}$, $[\hat{G}(\beta)]_{ij}=t_j(\beta) g_0(|{\bf r}_i-{\bf r}_j|)$, $[{\bm \psi}]_{i}={ \psi}({\bf r}_i)$, $[{\bm \phi}]_{i}={ \phi}({\bf r}_i)$, $i={1,...,N_s}$, from which ${\bm \psi}$ can be easily obtained, which completely determines the solution within the plate comprising scatterers and subjected to a point load.  
\begin{acknowledgments}
The presented work has been initiated under an Empa Internal Research Call scheme as project number 5213.00171.100.01. The
authors gratefully acknowledge the funding that made this work possible. The authors extend their gratitude to the colleagues Gwenael Hannema and Sven Vallely for their useful Ansys tips. 
\end{acknowledgments}

\bibliography{biblio}
\end{document}